\documentclass[manuscript,screen,acmsmall]{acmart}
\pdfoutput=1  

\newcommand{\markedchange}[1]{#1}

\setcopyright{acmlicensed}
\acmJournal{PACMHCI}
\acmYear{2022} \acmVolume{6} \acmNumber{CSCW2} \acmArticle{273} \acmMonth{11} \acmPrice{15.00}\acmDOI{10.1145/3555164}

\usepackage{amsmath}
\usepackage{tikz}
\usetikzlibrary{arrows,positioning}

\begin{document}

\title{Alexa as an Active Listener: How Backchanneling Can Elicit Self-Disclosure and Promote User Experience}

\author{Eugene Cho}
\email{choe@tcnj.edu}
\affiliation{%
  \institution{The College of New Jersey}
  \city{Ewing}
  \state{NJ}
  \country{USA}
}
\authornotemark[1]

\author{Nasim Motalebi}
\email{nfm5140@psu.edu}
\affiliation{%
  \institution{The Pennsylvania State University}
  \city{State College}
  \state{PA}
  \country{USA}
}
\authornote{Both authors contributed equally to the paper.}

\author{S. Shyam Sundar}
\email{sss12@psu.edu}
\affiliation{%
  \institution{The Pennsylvania State University}
  \city{State College}
  \state{PA}
  \country{USA}
}

\author{Saeed Abdullah}
\email{saeed@psu.edu}
\affiliation{%
  \institution{The Pennsylvania State University}
  \city{State College}
  \state{PA}
  \country{USA}
}

\renewcommand{\shortauthors}{Eugene Cho et al.}
\renewcommand{\shorttitle}{Alexa as an Active Listener}
\received{April 2021}
\received[revised]{November 2021}
\received[accepted]{March 2022}

\begin{abstract}
Active listening is a well-known skill applied in human communication to build intimacy and elicit self-disclosure to support a wide variety of cooperative tasks. When applied to conversational UIs, active listening from machines can also elicit greater self-disclosure by signaling to the users that they are being heard, which can have positive outcomes. However, it takes considerable engineering effort and training to embed active listening skills in machines at scale, given the need to personalize active-listening cues to individual users and their specific utterances. A more generic solution is needed given the increasing use of conversational agents, especially by the growing number of socially isolated individuals. With this in mind, we developed an Amazon Alexa skill that provides privacy-preserving and pseudo-random backchanneling to indicate active listening. User study (N = 40) data show that backchanneling improves perceived degree of active listening by smart speakers. It also results in more emotional disclosure, with participants using more positive words. Perception of smart speakers as active listeners is positively associated with perceived emotional support. Interview data corroborate the feasibility of using smart speakers to provide emotional support. These findings have important implications for smart speaker interaction design in several domains of cooperative work and social computing.
\end{abstract}

\ccsdesc[500]{Human-centered computing~Interaction design theory, concepts and paradigms}
\ccsdesc[500]{Applied computing~Consumer health}
\ccsdesc[300]{Human-centered computing~Personal digital assistants}

\keywords{smart speaker, voice user interface, attentive listening, eHealth}

\maketitle

\section{Introduction}
Active listening is a much-touted communication technique to convey one’s engagement in a given interaction, stimulate turn-taking and promote self-disclosure \cite{jonesSupportiveListening2011,wegerRelativeEffectivenessActive2014,hutchbyActiveListeningFormulations2005,millerOpenersIndividualsWho1983a,kuhn_power_2018}. While there are many similar techniques and alternative terms, including empathic, attentive and supportive listening, the essence of active listening comes from exhibiting unconditional attention to the speakers and confirmation of their experiences by utilizing certain responses such as verbal and nonverbal acknowledgment, paraphrasing, and asking disclosure-enhancing questions \cite{wegerRelativeEffectivenessActive2014}. By demonstrating various active listening skills, from generic (e.g., nodding and vocalizations) to specific (e.g., wincing or exclaiming), listeners serve as co-narrators by encouraging and influencing story telling from narrators \cite{conarrators}. 

Active listening by itself can be a form of therapy and support in both informal \cite{bodie_role_2015, kuhn_power_2018} and formal clinical settings
\cite{kornhaber_enhancing_2016,jagosh_importance_2011}. One mechanism by which active listening can help cope with stress and emotional episodes is through self-disclosure \cite{hutchbyActiveListeningFormulations2005,millerOpenersIndividualsWho1983a}. Active listening is found to effectively elicit disclosure of feelings and personal thoughts \cite{hutchbyActiveListeningFormulations2005,millerOpenersIndividualsWho1983a}, and such ``social sharing of emotion'' \cite{rime1991beyond} can lead to emotional recovery and distress reduction \cite{rime_social_1998}, thereby improving psychological wellbeing \cite{nils2012beyond}.

However, for active listening to be effective, it requires more than simply paying attention: active listening requires considerable effort and training \cite{robertson_active_2005}. While informal connections (e.g., family members) can provide support for self-disclosure, they might not have the necessary and consistent active listening skills to maximize emotional and psychological benefits \cite{kuhn_power_2018}. Furthermore, exposure to others’ emotional episodes can potentially impact listeners’ own wellbeing \cite{rauvola_compassion_2019,newell2010professional}. These issues can make it difficult for individuals in need to find adequate support to engage in self-disclosure behaviors in their day-to-day lives, which is particularly true for socially isolated individuals.

As an alternative to human listeners, we can leverage smart speakers with voice-based user interface (VUI) to support individuals to engage in effective disclosure in the real world. Smart speakers have seen a significant recent rise in adoption and ownership. \markedchange{In 2021, 94 million individuals in the US were estimated to own at least one smart speaker \cite{alcantaraSmart2021}}. The voice interfaces in these devices allow users to conduct a wide range of tasks. For example, users can perform search queries, get weather updates, and control appliances by ``talking'' to these devices. This project aims to extend the current capabilities of smart speakers to act as active listeners that can support self-disclosure.

Then, how can we make smart speakers to become active listeners? Active listening requires providing acknowledgment and avoiding interruptions to encourage individuals to elaborate and engage in disclosure over multi-turn dialogues \cite{jones2004putting}. Verbal continuers or backchanneling cues (e.g., ``hm'', ``ahm'', and ``yes'') can convey acknowledgment during conversations and thus, is an important element of active listening \cite{jones2004putting}. A timely backchanneling cue coming from a dialogue system can be enough to render users into thinking that the automated backchanneling responses came from a human counterpart \cite{Ward1996UsingPC}. Backchanneling cues have also been shown to support narrative development during a conversation with a spoken dialog system \cite{kawahara2016prediction,oertel2016towards}. \markedchange{Specifically in the context of smart-speaker interactions, Cohn et al. \cite{cohn-etal-2019-large} reported that systematically providing backchanneling cues to specific responses from an Alexa chatbot (i.e., fillers, such as "um", "mhmm") as well as emotionally expressive interjections that signal interest (e.g., "Awesome", "Cool") and express acceptance and agreement (e.g., "Okey dokey!", "High Five!") can enhance overall user rating of the interaction. This suggests that we can incorporate backchanneling cues in smart speakers to indicate active listening, which can subsequently lead to better support for user interactions, including self-disclosure}.

Yet, designing smart speakers to deliver backchanneling cues poses a number of technical challenges. Current smart speakers do not provide meaningful backchanneling cues to support narrative development, which is essential for effective disclosure. Instead, smart speakers are optimized for quick turn-taking and short verbal commands. Backchanneling also depends on context, flexible turn-taking, and timing of utterances. For instance, randomizing the backchanneling cues and switching backchanneling types (interchanging nodding and vocalized backchanneling) can reduce the human-likeness of an embodied virtual agent (compared to adhering to a human-created backchanneling cadence and type) \cite{TimingMatters}. However, offering context-aware backchanneling requires interpreting social cues and understanding natural language in real time, which can be highly challenging, especially given the absence of human-like embodied cues in smart speakers. 

Furthermore, computational limitations of smart speakers require offloading data processing steps to the cloud. This can raise serious privacy concerns in certain use cases, specifically when individuals might engage in potentially sensitive self-disclosure. For example, Lala et al. \cite{lala2017attentive} proposed the use of prosodic and content (e.g., focus word) features to deliver backchanneling. However, uploading and processing such features from self-disclosure in the cloud can result in considerable privacy, legal, and ethical challenges. As such, the key issue that we need to address before we can use smart speakers to support self-disclosure is this: \emph{how can we transform smart speakers into active listeners by providing verbal acknowledgments while minimizing privacy risks}?

We address this issue by delivering pseudo-random backchanneling instead of providing cues that are contingent upon the content of users’ commands and disclosures. To this end, we developed an Amazon Alexa app (``skill'') that delivers random backchanneling cues at pre-determined intervals as users engage with it. This system does not aim to be context-specific and as such, there is no need to store or process self-disclosure content. This considerably lowers privacy and security risks. We conducted a user study (N = 40) to evaluate the developed system. In the following sections, we describe the hypotheses guiding the study, as well as details of the methods used and findings emerging from statistical analyses of quantitative data and thematic analysis of qualitative data. We then discuss the implications of our findings to designing more privacy-sensitive and trustworthy VUIs and the applications of smart speaker technologies for therapeutic support. This study can help to extend the discussion within the CSCW community on social and collaborative user interactions with smart speakers, beyond supporting group interaction among users \cite{porcheron2017talking,porcheron2017animals}, by examining the potential of human-to-smart-speaker collaboration in support of both individuals’ and others’ (e.g., family members, caretakers, counselors) wellbeing. This study also expands previous CSCW studies focused on the use of conversational agents for stress-related disclosure \cite{sannon2018personification} and self-guided stress coping \cite{kamita2020promotion}.
\section{Background and Related Work}
Active listening has been studied as a major component in counseling \cite{danby2009listeners}, as a therapeutic skill that makes clients feel heard by the correspondent listening attentively and responding empathically \cite{levittcounsel}. It conveys immediate processing of information and decision making \cite{rost2014listening}. While there are several active listening tactics such as paraphrasing and asking questions \cite{wegerRelativeEffectivenessActive2014}, even the simplest verbal cues like backchanneling can have a significant impact on how others respond. \markedchange{In particular, Gardner \cite{gardner2001listeners} explained that such seemingly meaningless verbal cues actually imbue subtle and complex meanings into conversations as they help to co-create the narrative with other speakers}. In the context of telephone counseling, Danby et al. \cite{danby2009listeners} studied the role of backchanneling where counselors used minimal responses in the form of verbal acknowledgments and continuers (e.g., okay, mm hm, right) to encourage clients to talk. This listening behavior resulted in increased turn taking and decreased ambiguity compared to web-counseling through instant messaging \cite{danby2009listeners}. Such elicitation of disclosure is in fact a very important element of active listening that helps individuals cope with stress and emotional distress \cite{hutchbyActiveListeningFormulations2005,millerOpenersIndividualsWho1983a}.

Sharing of emotional experiences is a common practice across different populations and cultures \cite{rime2009emotion}. Such disclosure --- writing or talking about upsetting events --- can lead to psychological and physical health improvements \cite{frattaroli2006experimental}. Recent studies have proposed a number of theoretical models to explain positive effects of disclosure on health and wellbeing. For example, Sloan and Marx \cite{sloan2004taking} discussed how cognitive adaptation, disinhibition, and repeated exposure enabled by disclosure can lead to better processing of emotional events. Rimé \cite{rime2009emotion} also pointed out the socio-affective benefit of disclosure. Irrespective of the underlying mechanism, disclosure has been widely and successfully used as an intervention tool. Focusing on written disclosure, Pennebaker \cite{pennebaker1997writing} introduced expressive writing as a therapeutic process to address trauma. He found that increased use of insight, causal, and cognitive words in written disclosure is linked to health improvements. \markedchange{Expressive writing has been used to provide emotional support for individuals with mood disorder \cite{baikie2012expressive}, depression \cite{krpan2013everyday}, post-traumatic stress disorder (PTSD) \cite{bugg2009randomised}, and breast cancer \cite{craft2013expressive}}.

While recent disclosure studies have mostly focused on writing, verbal disclosure has also been shown to improve health and wellbeing \cite{nils2012beyond}. Indeed, Fattarolli \cite{frattaroli2006experimental} argued that speaking may be superior to writing as it is easier and demands fewer cognitive resources. Furthermore, speaking might also allow more complex disclosure through both verbal and non-verbal (e.g., facial) expressions. The interactive nature of verbal disclosure requires two actors --- a major speaker and a listener --- to construct the narrative and maintain engagement. The outcome of verbal disclosure depends on listeners’ reactions \cite{rime2009emotion} --- active listening leads to effective disclosure. 

A number of recent studies have focused on developing technologies that can increase engagement by serving as active listeners. \markedchange{Johansson et al. \cite{johansson2016making} developed a social robot that incorporates listening cues in its responses when asking users to disclose their travel memories}. Their study explores the process of developing a dialogue system and a response model that allows a robot to act as an active listener. Similarly, DeVault \cite{devault2011incremental} developed an embodied virtual agent called Semi Sensei Kiosk that demonstrates empathic listening behavior using verbal continuers to engage in small talk with patients who have PTSD or depression. Lala et al. \cite{lala2017attentive} developed Erica --- an embodied bot for the elderly to maintain communication abilities. \markedchange{Their system not only uses backchanneling to show interest and engage with the user, but also produces context-specific responses and aims to support appropriate turn-taking behaviors}.

Such development efforts stem from the fact that technologies can act as effective agents to elicit self-disclosure from users. For instance, Ho et al. \cite{ho2018psychological} compared the effects of written emotional disclosure to a chatbot vs. a human partner, and found that both produce similar psychological benefits. Regardless of agent (i.e., chatbot vs. human), users were found to make more intimate disclosures when they engaged in emotional (vs. factual) conversations, which in turn, created positive emotional and relational outcomes \cite{ho2018psychological}. It seems that adding certain interaction cues based on human communication principles can encourage users’ self-disclosure to machines. Relevant to this point, Moon \cite{10.1086/209566} found that when computers revealed intimate information first and followed socially appropriate sequence of disclosure (i.e., gradual disclosure from superficial to intimate), users became more comfortable making intimate self-disclosure, just as in human-to-human communication. Moving from computers, Lee et al. \cite{lee2020hear} found that reciprocal self-disclosure from chatbots also elicited more disclosure from users. Liu and Sundar \cite{liu2018should} reported that when a chatbot expressed sympathy or empathy, it was perceived as being more understanding and supportive. Jeong et al. \cite{jeong2019exploring} found that the use of ``conversational fillers'' by conversational agents made them more entertaining for socially oriented interactions.

Together, this body of research suggests active listening by non-human agents or bots could indeed be psychologically significant, affecting disclosure and subsequent emotional state. Extending those findings into the context of smart speakers, we hypothesize that active listening signaled by backchanneling cues coming from the smart speaker can exert positive effects on three aspects of emotional and behavioral outcomes: (a) change in emotional states, (b) perceived emotional support, and (c) self-disclosure behaviors (H2). \markedchange{However, for such effects to occur, a prerequisite condition is that users need to recognize a smart speaker with backchanneling cues as an active listener (H1)}. Nass and Lee \cite{nasslee2001} label such recognition (e.g., identification and classification) of social characteristics of computers (or non-human objects) as "first-degree social response", triggering which can lead to more applied attitudinal and behavioral changes among users (i.e., "second-degree social response"). Following this rationale, we test both first-degree (H1) and second-degree (H2) social responses from users, focused on smart speaker backchanneling effects. In addition, we examine this cascading effect of recognition (first-degree) to application (second-degree) regarding users' socialization with smart speakers by testing the mediating role of active listening perception in the backchanneling effect on user outcomes (H3). In doing so, this study can help shed light on one of the underlying theoretical mechanisms that explains users' social responses to smart speakers. \markedchange{Overall, this paper focuses on the following hypotheses}:

\begin{itemize}
    \item{\textbf{H1}}: Backchanneling cues will positively affect users’ perception of a smart speaker as an active listener (i.e., active listening perception).

    \item{\textbf{H2}}: Backchanneling cues will have positive effects on users’ (a) emotional state, (b) perceived emotional support, and (c) self-disclosure behaviors.

    \item{\textbf{H3}}: Active listening perception will mediate the effects of backchanneling cues on positive outcomes predicted by H2.

\end{itemize}

Furthermore, beyond promoting emotional support and self-disclosure, we explored if the above outcomes will eventually lead to a better user experience. In particular, by incorporating pseudo-random backchanneling cues, we not only test if backchanneling from smart speakers can effectively deliver active-listening benefits, but also if it can be provided in a privacy-preserving yet user-friendly manner, leading to the following research question:

\begin{itemize}
    \item{\textbf{RQ1}}: Will the hypothesized positive effects of backchanneling eventually lead to improved usability of the smart speakers?
\end{itemize}
    
\section{Method}

To determine the effects of pseudo-random backchanneling, this study employed a
between-subjects experiment design with 2 conditions --- backchanneling vs.
control. Participants in these conditions interacted with different Alexa
skills. They were not aware of the difference between the control and backchanneling conditions assigned to them, nor aware of the purpose of our study. During the study, we collected audio recordings of participants'
interactions with Alexa using a separate audio recorder. We recruited 40 participants from a large public
university in the United States and assigned them to one of the two conditions.
The gender distribution of the participants was even with 20 males and 20
females \markedchange{(see Table \ref{table:participants} for individual participant information)}. Only half of the participants mentioned that English was their first language, but all participants were able to understand the study procedure and interact with Alexa fluently in English. Of the 40 participants, 24 participants (60\%)
stated that they currently use smart speaker assistants (14 infrequent users and 10 regular users), and 15 users reported to not use smart speaker assistant at the time of data collection.

\begin{table}[htb]
\begin{tabular}{||c c c c c c||} 
 \hline
 ID & Age & Gender & Native in English & Education & Smart Speaker Use \\ [0.5ex]
 \hline\hline
 P1 & 21-24 & Female & Yes & College Student & Yes, regularly \\
 \hline
 P2 & 21-24 & Female & No & College Student & Yes, regularly \\
 \hline
 P3 & 18-20 & Male & Yes & College Student & No \\
 \hline
 P4 & 18-20 & Male & Yes & College Student & No \\
 \hline
 P5 & 31-34 & Female & No & Graduate Student & Yes, regularly \\
 \hline
 P6 & 18-20 & Male & Yes & College Student & Yes, but not often \\
 \hline
 P7 & 35-40 & Female & Yes & College Alumna & Yes, but not often \\
 \hline
 P8 & 25-30 & Male & No & Graduate Student & Yes, regularly \\
 \hline
 P9 & 25-30 & Female & No & Post-graduate & Yes, regularly \\
 \hline
 P10 & 25-30 & Male & Yes & Graduate Student & No \\
 \hline
 P11 & 25-30 & Male & No & Graduate Student & Yes, but not often \\
 \hline
 P12 & 31-34 & Male & No & Graduate Student & Yes, regularly \\
 \hline
 P13 & 31-34 & Female & No & College Alumna & Yes, regularly \\
 \hline
 P14 & 25-30 & Male & No & Graduate Student & Yes, regularly \\
 \hline
 P15 & 21-24 & Male & No & College Student & Yes, but not often \\
 \hline
 P16 & 25-30 & Female & No & Graduate Student & No \\
 \hline
 P17 & 31-34 & Male & Yes & Post-graduate & Yes, but not often \\
 \hline
 P18 & 25-30 & Male & No & Graduate Student & Yes, but not often \\
 \hline
 P19 & 25-30 & Female & No & Graduate Student & Yes, but not often \\
 \hline
 P20 & Older than 50 & Female & Yes & College Alumna & Yes, regularly \\
 \hline
 P21 & 31-34 & Female & No & Graduate Student & No \\
 \hline
 P22 & 18-20 & Female & Yes & College Student & Yes, but not often \\
 \hline
 P23 & Older than 50 & Female & Yes & Post-graduate & Yes, regularly \\
 \hline
 P24 & 25-30 & Male & Yes & Graduate Student & Yes, but not often \\
 \hline
 P25 & 25-30 & Female & Yes & Graduate Student & No \\
 \hline
 P26 & 25-30 & Female & No & College Student & Yes, but not often \\
 \hline
 P27 & 25-30 & Male & No & Graduate Student & Yes, regularly \\
 \hline
 P28 & 21-24 & Male & Yes & Graduate Student & Yes, but not often \\
 \hline
 P29 & 18-20 & Female & Yes & College Student & Yes, regularly \\
 \hline
 P30 & 21-24 & Female & No & Graduate Student & No \\
 \hline
 P31 & 25-30 & Male & No & Graduate Student & Yes, but not often \\
 \hline
 P32 & 25-30 & Male & No & Graduate Student & No \\
 \hline
 P33 & 25-30 & Male & No & Graduate Student & Yes, but not often \\
 \hline
 P34 & 25-30 & Male & No & Graduate Student & Yes, but not often \\
 \hline
 P35 & 25-30 & Female & Yes & College Student & Yes, regularly \\
 \hline
 P36 & 21-24 & Female & Yes & College Alumna & Yes, regularly \\
 \hline
 P37 & 25-30 & Female & Yes & College Alumna & Yes, regularly \\
 \hline
 P38 & 35-40 & Male & Yes & Post-graduate & No \\
 \hline
 P39 & 25-30 & Male & Yes & Graduate Student & No \\
 \hline
 P40 & Older than 50 & Female & Yes & College Alumna & Yes, regularly \\
 \hline
\end{tabular}
\caption{\label{table:participants}Participant information}
\end{table}

\subsection{Procedures}
During the study, participants first provided consent and then rated their
current emotional state through an online questionnaire, which served as our
pre-test measure. After that, they were instructed to interact with Alexa to
express their thoughts regarding certain personal life matters. The skills asked
users to answer 4 questions: two questions related to professional life regarding time management and
weekly goals, and two questions regarding personal life choices and
relationships (see the supplementary document for the questions used for the user interaction with Alexa). We incorporated
specific questions in the study since our pilot data showed that some
participants have difficulty in coming up with topics to interact with Alexa.

To control for the order effect of question type, half the participants in
each condition were asked the professional questions first and the other half were
asked the personal life and relationship questions first. Participants were not interrupted by the
researcher during the study. To provide a sense of privacy, the researcher left
the room as participants interacted with Alexa. Since the Alexa skill we used for this study did not record any of the user comments (as described in the next section), we used a separate device to
record users' interactions with Alexa, which were manually transcribed later for text
analysis.

After their interactions with Alexa, participants completed an online
questionnaire with questions related to the outcome variables of interest and
demographic information. In addition, they went through a semi-structured interview
regarding their experiences, expectations, and ideas for the future of smart
speakers for mental health. Upon completion, participants were compensated \$10
for their time. The research protocol was approved by the Institutional
Review Board (IRB) from the researchers' institution at the time of data collection. All of the participants were informed that the entire study session will be recorded for analysis, and offered consent prior to participation. We did not physically disable the microphone as the rest of study session required functional interactions with Alexa. However, the skill did not collect any responses as the participants were speaking about their personal experiences.

\subsection{Manipulation and Alexa Skill Development}

For the manipulation of the independent variable (i.e., active listening vs.
control), we developed two different sets of Alexa skills. While skills for both
conditions asked users to respond to four questions as mentioned above, only the
active listening condition provided pseudo-random verbal continuers such as
``hm,'' ``yeah,'' ``go on,'' and ``I see''. No such cues were added to the Alexa skill developed for the control condition.

When developing the Alexa skill, we initially leveraged the Amazon Alexa framework
to deliver context-specific backchanneling cues to users. For example, we
conducted sentiment analysis to deliver appropriate continuers (e.g., responding
``I am sorry to hear that'' to a negative event mentioned by the participant),
and developed a number of prototypes and conducted evaluation over time.
However, given the current limitations of Alexa API, delivering fully responsive
and context-specific backchanneling cues turned out to be unfeasible.
Specifically, the Amazon Alexa ecosystem is optimized for short sentences. The
devices also have limited processing capabilities and offload bulk of their data
handling to the cloud. This causes significant delay in processing
multi-sentence user inputs. In other words, network and processing delays make
it impractical to deliver dynamic backchanneling cues in a time-appropriate
manner. Other smart speaker platforms have similar limitations.
Furthermore, context-specific backchanneling requires speech content processing,
which can have serious privacy issues. Given these limitations, we instead
focused on a context-independent way of providing backchanneling cues. That is,
instead of performing prosodic and content feature analysis to identify end of
utterance and context, we decided to deliver pseudo-random backchanneling cues
at pre-determined intervals. Thus, we note that the Alexa skill we developed and used for this study were not able to listen, understand, nor detect user speech content.

\begin{figure}[htb]
\begin{center}
    \includegraphics[width=0.95\textwidth]{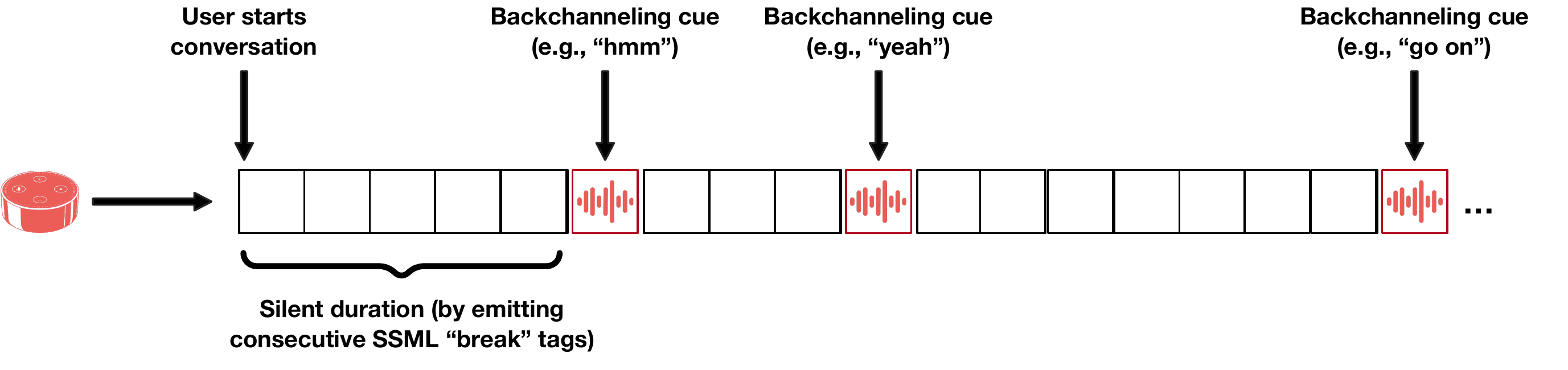}

    \caption{\label{fig:alexa-framework}We used Speech Synthesis Markup Language (SSML) \cite{amazoncom_inc_speech_nodate} to embed backchanneling cues during user interaction for the active listening condition.}

    \Description{Schematic diagram shows the Alexa skill interaction. When user
    starts conversation, the skill  delivers silent durations (by emitting
    consecutive SSML ``break'' tags). The skill provides pseudo-random
    backchanneling audio cues (e.g., hmm, yeah, go on).}

\end{center}
\end{figure}

For implementation, we used the Amazon Alexa framework (see Figure
\ref{fig:alexa-framework}). For generating pauses (silent durations), we used
the ``break'' tag in Speech Synthesis Markup Language (SSML)
\cite{amazoncom_inc_speech_nodate} supported by Alexa. The duration of the
pauses ranged between 7 to 50 seconds. These pause durations were selected based on pilot data \cite{motalebi2019} to minimize user interruptions. Given that a ``break'' tag can have 10
seconds of silence at most, we incorporated multiple ``break'' tags whenever
necessary. After a pause, Alexa delivered a backchanneling cue. We ensured that all
consecutive backchanneling cues were different. To avoid interrupting users, we
also used lower volume for backchanneling cues (by using the parameter ``soft''
for volume in the ``prosody'' SSML tag). During the study session, we did not
record any user responses using Alexa. This allowed us to avoid potential
privacy concerns.

\subsection{Measurements}

\subsubsection{Perception of active listening}

We adapted the Active-Empathic Listening scale from Gearhart and Bodie
\cite{gearhart2011active} to assess perceived active listening by Alexa
(e.g., ``Alexa was sensitive to what I was saying;'' ``Alexa seemed to listen
to me for more than just spoken words,'' ``Alexa delivered a sense of agreement
for what I was saying when appropriate;'' 10-point Likert scale; M = 3.23, SD =
1.93, $\alpha$ = .94).

\subsubsection{Emotional states}

To evaluate change in users' emotions after interacting with Alexa, we
administered two questionnaires: once before (pre-emotion) and another after
(post-emotion) the interaction with Alexa. Following Watson et al.
\cite{watson1988development}, participants were asked to rate their current
emotional state on the following 10 affective adjectives: active, calm, content,
enthusiastic, happy, angry, anxious, embarrassed, nervous, and sad.  Five of
these adjectives indicate positive emotions, and other five adjectives indicate
negative emotions. These ratings were measured on a 10-point Likert scale (see
Table \ref{table:measurements}).

\subsubsection{Perceived emotional support from Alexa}
We revised 6 items from Clark et al. \cite{clark1998impact}, originally designed
to measure the effectiveness of comforting messages from friends, to fit the
context of interaction with Alexa (e.g., ``Alexa made me feel better about
myself,'' ``After talking with the Alexa, I feel less depressed,'' ``Talking
with Alexa helped me get my mind off the negative experience'' 10-point Likert
scale; M = 5.12, SD = 2.10; $\alpha$ = .94).

\subsubsection{Users' self-disclosure behaviors}

Self-disclosure behaviors of users were measured in both the amount of
self-expression (by counting (1) time spent and (2) words used in interaction
with Alexa) as well as emotional expression (by identifying the proportion of
(3) positive and (4) negative emotional word usage). These behavioral measures
were obtained from the audio recordings of the interaction. In particular, the
length of the audio interaction represented (1) time spent on the interaction
(M = 346.49 seconds, SD = 153.18). Following Balon and Rim{\'e}
\cite{balon2016lexical}, we used LIWC (Linguistic Inquiry and Word Count)
\cite{tausczikPsychologicalMeaningWords2010} to perform lexical analysis. This
included calculating word count (M = 672.03 words, SD = 339.10), and the
proportion of the use of positive (M = 2.20\%, SD = 1.05) and negative (M =
0.55\%, SD = 0.44) emotional words. Due to technical issues, we were not able
to extract audio recording of one participant's interaction. We also asked
about demographics (e.g., language skills) and media usage patterns in our
questionnaire \footnote{We have uploaded the surveys and questionnaires used in
the study as supplementary documents.}.

\subsubsection{Perceived usability of the Alexa skill}

Participants were asked to evaluate the quality of the Alexa skill based on the
following 11 adjectives: good, useful, high quality, user-friendly, coherent,
organized, pleasant, entertaining, appealing, intelligent, smart; (M = 6.09, SD
= 2.18, $\alpha$ = .96). Those items were adapted from previous research to
measure attitudes toward web services related to usability
\cite{kalyanaraman2006psychological,sundar2011beyond}, with some items being
added to represent ``smartness'' of the Alexa skill. These items used a
10-point Likert scale. As supplementary usability measures for general
expectations and evaluations of Alexa's responses, we elicited participants'
agreement level with the following 5 questions on a 10-point Likert scale:
``The way Alexa talked to me irritated me,'' ``Alexa's responses were
appropriate,'' ``I felt like Alexa was putting me down,'' ``I wish Alexa's
responses had been briefer,'' and ``I wish Alexa's responses had been longer.''

\begin{table}[htb] \begin{tabular}{|c|c|c|c|} \hline \textbf{Emotions} &
\textbf{\begin{tabular}[c]{@{}c@{}}Pre-interaction\\ Mean (SD)\end{tabular}} &
\textbf{\begin{tabular}[c]{@{}c@{}}Post-interaction\\ Mean (SD)\end{tabular}} &
\textbf{\begin{tabular}[c]{@{}c@{}}Pre vs. Post\\ Difference\end{tabular}} \\
\hline Active            & 6.78 (2.21) & 6.70 (2.65) & F(1,38)=.07, p=.79
\\ \hline Calm              & 7.10 (2.31) & 7.10 (2.35) & F(1,38)=.001, p=1.0
\\ \hline Content           & 6.40 (2.70) & 6.67 (2.36) & F(1,38)=.71, p=.41
\\ \hline Enthusiastic      & 6.43 (2.17) & 6.25 (2.50) & F(1,38)=.37, p=.55
\\ \hline Happy             & 6.67 (2.30) & 6.77 (2.43) & F(1,38)=.11, p=.74
\\ \hline Angry             & 1.65 (1.41) & 1.68 (1.49) & F(1,38)=.03, p=.87
\\ \hline Anxious           & 2.73 (1.99) & 2.25 (1.95) & F(1,38)=4.51, p=.09
\\ \hline Embarrassed        & 1.75 (1.35) & 1.70 (1.49) & F(1,38)=.04, p=.85
\\ \hline Nervous           & 2.68 (2.21) & 2.10 (1.81) & F(1,38)=8.58, p $<$
.01                                                    \\ \hline Sad
& 2.73 (2.33) & 2.30 (2.10) & F(1,38)=4.06, p=.05
\\ \hline \end{tabular}
    \caption{\label{table:measurements}Pre- and post-interaction emotional
    states}
\end{table}

\section{Results}

Before the main analyses, we investigated whether self-disclosure behaviors
differ across demographic variables in our dataset.  While non-native speakers
tended to use less words (t(37) = 2.32, p = .03, $\eta_{p}^{2}$ = .13), and spend less time (t(37) =
1.20, p = .053, $\eta_{p}^{2}$ = .10) on the interaction compared to native speakers, their use of
positive and negative emotional words did not differ significantly (ps = .24).
Also, current smart-speaker users did not differ from non-users on any of the 4
self-disclosure behaviors (ps = .33). More importantly, the significant main
effects of backchanneling cues on active listening perceptions (see section
4.1. for specific findings) remained significant after controlling for language
difference as well as previous usage (F(1, 36) = 5.06, p = .03, $\eta_{p}^{2}$ = .12). In other
words, \textbf{self-disclosure behaviors were mostly consistent across
different demographic groups}.

\subsection{Effects of backchanneling on active listening perceptions}

We used an independent-samples t-test to assess whether the use of
backchanneling cues enhances users' perception of smart \markedchange{speakers as active
listeners} (H1). Participants in the backchanneling condition (M = 3.91, SD =
1.96) evaluated Alexa as being a better active listener compared to the
control condition (M = 2.56, SD = 1.68) with (t(38) = 2.34, p = .03, $\eta_{p}^{2}$ = .13). Therefore, \textbf{our data support H1}.

\subsection{Emotional contrast before and after interaction with Alexa}

We also examined if backchanneling cues had any effects on users' (a) change in
emotional state, (b) self-disclosure behaviors, and (c) perceived emotional
support, as suggested in H2. We used a 2 (backchanneling vs. control) $\times$
2 (pre- vs. post-emotions) mixed model repeated-measures analysis of variance
(ANOVA) to examine the effects on (a) change in emotional state. The findings
showed no significant differences between the backchanneling and control groups
in emotional changes during the interaction, for any of the 10 emotions (ps $>$
.12). As such, our data did not support H2a.  There was however a significant
reduction in nervousness (F(1,38) = 8.58, p = .006, $\eta_{p}^{2}$ = .18) after interacting with Alexa (see Table \ref{table:measurements}; due to the non-significant difference in emotional states between the backchanneling and control conditions, the emotional state of two groups are combined).

To further examine if backchanneling had an effect on (b) perceived emotional
support and (c) self-disclosure behaviors, we conducted a series of
independent-samples t-tests. We did not find evidence to suggest that backchanneling improved
perceived social support obtained from Alexa (H2b; t(38) = 0.31, p = .76, $\eta_{p}^{2}$ = .003). In
terms of self-disclosure behaviors (H2c), backchanneling was not associated with more
interaction with Alexa in both time spent (t(37) = 0.33, p = .75, $\eta_{p}^{2}$ = .003) and words
used (t(37) = 0.56, p = .58, $\eta_{p}^{2}$ = .01), nor promoted more use of positive emotional words
(t(37) = 1.15, p = .26, $\eta_{p}^{2}$ = .03).  However, backchanneling had a significant effect on the use of
negative words (t(37) = 2.26, p = .03, $\eta_{p}^{2}$ = .12) --- participants used more negative
words in the backchanneling condition (M = 0.70, SD = 0.47), compared to the
control (M = 0.40, SD = 0.35). \textbf{Overall, our data, thus, did not support H2}.

\tikzstyle{box}=[rectangle,draw,text width=3.2cm,text centered, font=\bfseries,thick]
\tikzstyle{small-box}=[rectangle,draw,text width=2.5cm,text centered, font=\bfseries,thick]
\tikzstyle{tiny-box}=[rectangle,draw,text width=2.0cm,text centered, font=\bfseries,thick]
\tikzstyle{arrow}=[shorten >=1pt,>=stealth',thick]


\begin{figure}[t]
\begin{center}
    \begin{tikzpicture}[auto, node distance=1.5cm]
        \node[box] (r1) at (0:0) {Backchanneling\\Condition (vs Control)};
        \node[box,right=of r1] (r2) {Active Listening\\Perception};
        \node[box,right=of r2] (r3) {Change in Enthusiastic\\Emotional State};
        
        \draw[->,arrow] (r1) to
                node[auto,below]{
                    $\begin{aligned}
                        b &= 1.35\\
                        t &= 2.34\\
                        p &= 0.02\\
                    \end{aligned}$
                }
            (r2);

        \draw[->,arrow,dashed] (r2) to
                node[auto,below]{
                    $\begin{aligned}
                        b &= 0.31\\
                        t &= 2.00\\
                        p &= 0.053\\
                    \end{aligned}$
                }
            (r3);

        \draw[->, arrow,dashed] (r1.south) -- ++(0, -1.5cm)
                node[below=1.7cm of r2]{
                    $\begin{aligned}
                        b &= -0.17\\
                        t &= -0.29\\
                        p &= 0.78\\
                    \end{aligned}$
                }
            -| (r3.south);

    \end{tikzpicture}
\caption{\label{fig:result-enthusiasm}Significant positive mediation effects of backchanneling on change in enthusiastic emotional state through increased active listening perception.}

    \Description{Effect of backchanneling on Active Listening Perception (b
    = 1.35, t = 2.34, p = 0.02). Effect of Active Listening Perception on
    Change in Enthusiastic Emotional State (b = 0.31, t = 2.00, p = 0.053).
    Indirect effect of backchanneling (b = −0.17, t = −0.29, p = 0.78)}

\end{center}
\end{figure}
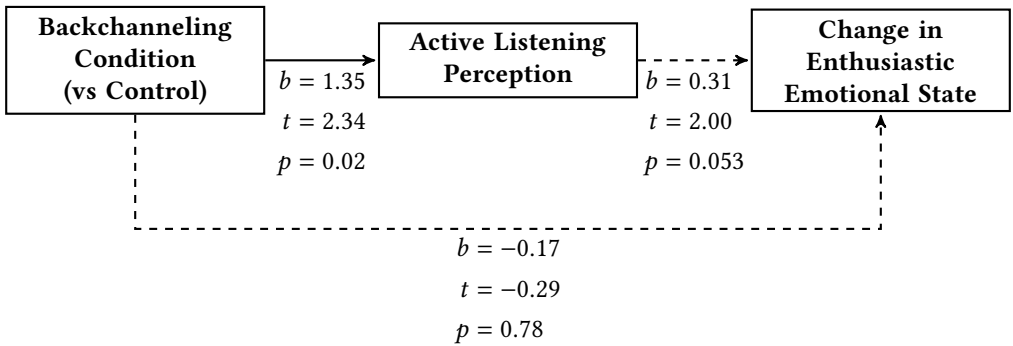


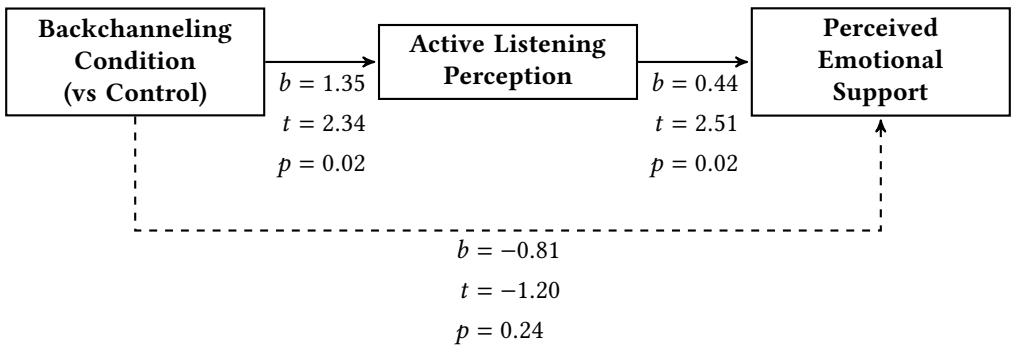
\begin{figure}[b]
\begin{center}
    \begin{tikzpicture}[auto, node distance=1.5cm]
        \node[box] (r1) at (0:0) {Backchanneling\\Condition (vs Control)};
        \node[box,right=of r1] (r2) {Active Listening\\Perception};
        \node[box,right=of r2] (r3) {Perceived Emotional\\Support};
        
        \draw[->,arrow] (r1) to
                node[auto,below]{
                    $\begin{aligned}
                        b &= 1.35\\
                        t &= 2.34\\
                        p &= 0.02\\
                    \end{aligned}$
                }
            (r2);

        \draw[->,arrow] (r2) to
                node[auto,below]{
                    $\begin{aligned}
                        b &= 0.44\\
                        t &= 2.51\\
                        p &= 0.02\\
                    \end{aligned}$
                }
            (r3);

        \draw[->, arrow,dashed] (r1.south) -- ++(0, -1.5cm)
                node[below=1.7cm of r2]{
                    $\begin{aligned}
                        b &= -0.81\\
                        t &= -1.20\\
                        p &= 0.24\\
                    \end{aligned}$
                }
            -| (r3.south);

    \end{tikzpicture}
\caption{\label{fig:result-perceived-support}Significant positive mediation effects of backchanneling on perceived emotional support from Alexa through increased active listening perception.}

    \Description{Effect of backchanneling on Active Listening Perception (b
    = 1.35, t = 2.34, p = 0.02). Effect of Active Listening Perception on
    Change in Enthusiastic Emotional State (b = 0.31, t = 2.00, p = 0.053).
    Indirect effect of backchanneling (b = −0.17, t = −0.29, p = 0.78)}

\end{center}
\end{figure}


\begin{figure}[t]
\begin{center}
    \begin{tikzpicture}[auto, node distance=1.5cm]
        \node[box] (r1) at (0:0) {Backchanneling\\Condition (vs Control)};
        \node[box,right=of r1] (r2) {Active Listening\\Perception};
        \node[box,right=of r2] (r3) {Use of Positive\\Emotional Words};
        
        \draw[->,arrow] (r1) to
                node[auto,below]{
                    $\begin{aligned}
                        b &= 1.50\\
                        t &= 2.62\\
                        p &= 0.01\\
                    \end{aligned}$
                }
            (r2);

        \draw[->,arrow] (r2) to
                node[auto,below]{
                    $\begin{aligned}
                        b &= 0.19\\
                        t &= 2.10\\
                        p &= 0.04\\
                    \end{aligned}$
                }
            (r3);

        \draw[->, arrow,dashed] (r1.south) -- ++(0, -1.5cm)
                node[below=1.7cm of r2]{
                    $\begin{aligned}
                        b &= 0.09\\
                        t &= 0.27\\
                        p &= 0.79\\
                    \end{aligned}$
                }
            -| (r3.south);

    \end{tikzpicture}
\caption{\label{fig:result-positive-words}Significant positive mediation effects of backchanneling on the use of positive emotional words through increased active listening perception.}
    
    \Description{Effect of backchanneling on Active Listening Perception (b
    = 1.35, t = 2.34, p = 0.02). Effect of Active Listening Perception on
    Perceived Emotional Support (b = 0.44, t = 2.51, p = 0.02). Indirect effect
    of backchanneling on Use of Positive Emotional Words (b = −0.81, t = −1.20,
    p = 0.24)}

\end{center}
\end{figure}
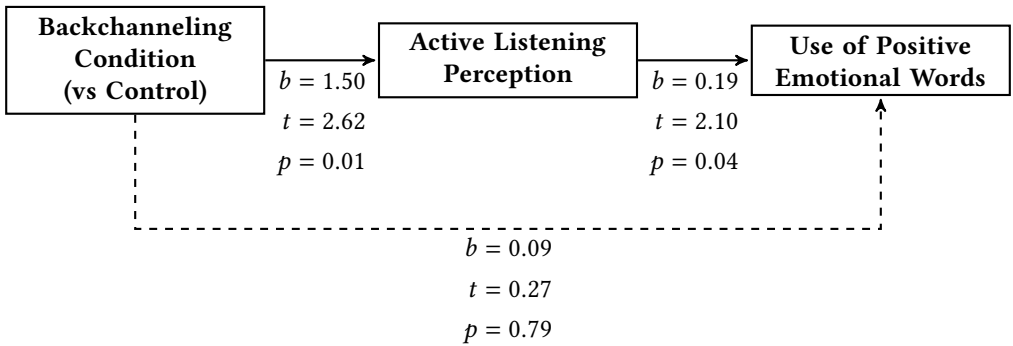


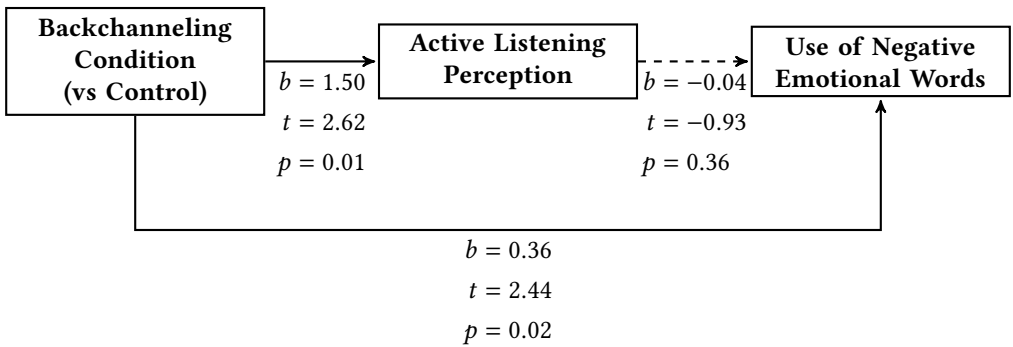
\begin{figure}[b]
\begin{center}
    \begin{tikzpicture}[auto, node distance=1.5cm]
        \node[box] (r1) at (0:0) {Backchanneling\\Condition (vs Control)};
        \node[box,right=of r1] (r2) {Active Listening\\Perception};
        \node[box,right=of r2] (r3) {Use of Negative\\Emotional Words};
        
        \draw[->,arrow] (r1) to
                node[auto,below]{
                    $\begin{aligned}
                        b &= 1.50\\
                        t &= 2.62\\
                        p &= 0.01\\
                    \end{aligned}$
                }
            (r2);

        \draw[->,arrow,dashed] (r2) to
                node[auto,below]{
                    $\begin{aligned}
                        b &= -0.04\\
                        t &= -0.93\\
                        p &= 0.36\\
                    \end{aligned}$
                }
            (r3);

        \draw[->, arrow] (r1.south) -- ++(0, -1.5cm)
                node[below=1.7cm of r2]{
                    $\begin{aligned}
                        b &= 0.36\\
                        t &= 2.44\\
                        p &= 0.02\\
                    \end{aligned}$
                }
            -| (r3.south);

    \end{tikzpicture}
\caption{\label{fig:result-negative-words} Non-significant mediation effects of backchanneling on the use of negative emotional words through active listening perception.}

    \Description{Effect of backchanneling on Active Listening Perception (b
    = 1.50, t = 2.62, p = 0.01). Effect of Active Listening Perception on
    Use of Negative Emotional Words (b = −0.04, t = −0.93, p = 0.36). Indirect
    effect of backchanneling on Use of Negative Emotional Words (b = 0.36, t =
    2.44, p = 0.02)}

\end{center}
\end{figure}


\begin{figure}
\begin{center}
    \begin{tikzpicture}[auto, node distance=1.3cm]
        \node[small-box] (r1) at (0:0) {Backchanneling\\Condition (vs Control)};
        \node[tiny-box,right=of r1] (r2) {Active\\Listening\\Perception};
        \node[tiny-box,right=of r2] (r3) {Perceived Emotional\\Support};
        \node[tiny-box,right=of r3] (r4) {Perceived Usability\\of Alexa};
        
        \draw[->,arrow] (r1) to
                node[auto,below]{
                    $\begin{aligned}
                        b &= 1.35\\
                        t &= 2.34\\
                        p &= 0.02\\
                    \end{aligned}$
                }
            (r2);

        \draw[->,arrow] (r2) to
                node[auto,below] (midlabel) {
                    $\begin{aligned}
                        b &= 0.44\\
                        t &= 2.51\\
                        p &= 0.02\\
                    \end{aligned}$
                }
            (r3);

        \draw[->,arrow] (r3) to
                node[auto,below]{
                    $\begin{aligned}
                        b &= 0.58\\
                        t &= 4.92\\
                        p &< 0.001\\
                    \end{aligned}$
                }
            (r4);

        \draw[->, arrow, dashed] (r1.south) -- ++(0, -1.5cm)
                node[below=.6cm of midlabel]{
                    $\begin{aligned}
                        b &= -0.51\\
                        t &= -1.04\\
                        p &= 0.31\\
                    \end{aligned}$
                }
            -| (r4.south);

    \end{tikzpicture}
\caption{\label{fig:result-support-usability}Significant positive serial indirect effects of backchanneling on perceived usability through increased active listening perception and emotional support.}

    \Description{Effect of backchanneling on Active Listening Perception (b
    = 1.35, t = 2.34, p = 0.02). Effect of Active Listening on Perceived
    Emotional Support (b =  0.44, t = 2.51,  p = 0.02).  Effect of Perceived
    Emotional Support on Perceived Usability of Alexa (b = 0.58, t = 4.92, p
    $<$ 0.001).  Indirect effect of backchanneling on Perceived Usability of
    Alexa (b = −0.51, t = −1.04, p = 0.31)}

\end{center}
\end{figure}
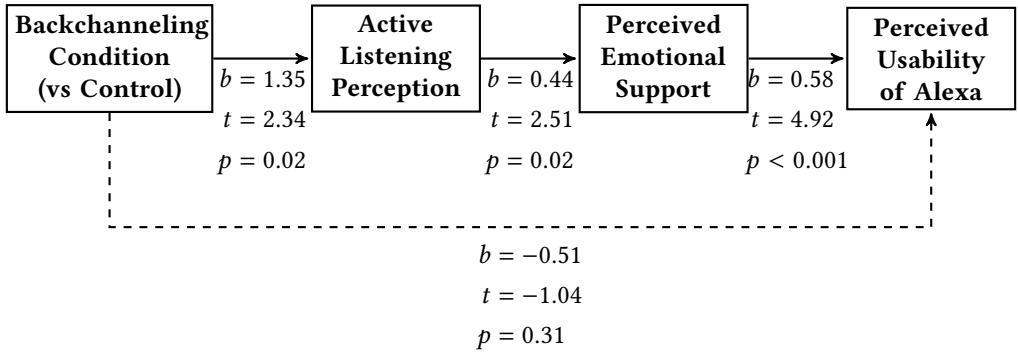


\begin{figure}
\begin{center}
    \begin{tikzpicture}[auto, node distance=1.4cm]
        \node[small-box] (r1) at (0:0) {Backchanneling\\Condition (vs Control)};
        \node[tiny-box,right=of r1] (r2) {Active\\Listening\\Perception};
        \node[tiny-box,right=of r2] (r3) {Positive Emotional\\ Words};
        \node[tiny-box,right=of r3] (r4) {Perceived Usability\\of Alexa};
        
        \draw[->,arrow] (r1) to
                node[auto,below]{
                    $\begin{aligned}
                        b &= 1.50\\
                        t &= 2.62\\
                        p &= 0.01\\
                    \end{aligned}$
                }
            (r2);

        \draw[->,arrow] (r2) to
                node[auto,below] (midlabel) {
                    $\begin{aligned}
                        b &= 0.19\\
                        t &= 2.10\\
                        p &= 0.04\\
                    \end{aligned}$
                }
            (r3);

        \draw[->,arrow,dashed] (r3) to
                node[auto,below]{
                    $\begin{aligned}
                        b &= 0.08\\
                        t &= 0.24\\
                        p &= 0.81\\
                    \end{aligned}$
                }
            (r4);

        \draw[->, arrow, dashed] (r1.south) -- ++(0, -1.5cm)
                node[below=.6cm of midlabel]{
                    $\begin{aligned}
                        b &= -0.95\\
                        t &= -1.46\\
                        p &= 0.15\\
                    \end{aligned}$
                }
            -| (r4.south);

    \end{tikzpicture}
\caption{\label{fig:result-words-usability}Non-significant serial indirect effects of backchanneling on perceived usability through active listening perception and the use of positive emotional words.}

    \Description{Effect of backchanneling on active listening Perception (b
    = 1.50, t = 2.62, p = 0.01). Effect of active listening on Positive
    Emotional Words (b =  0.19, t = 2.10,  p = 0.04).  Effect of Positive
    Emotional Words on Perceived Usability of Alexa (b = 0.08, t = 0.24, p =
    0.81).  Indirect effect of backchanneling on Perceived Usability of Alexa
    (b = −0.95, t = −1.46, p = 0.15)}

\end{center}
\end{figure}
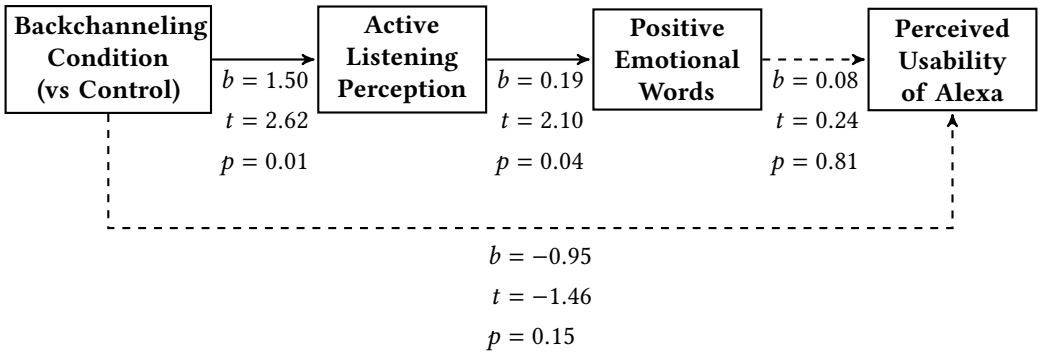

\subsection{Mediation effects toward emotional support and self-disclosure}

As proposed in H3, it is possible that backchanneling can indirectly improve
emotional feelings and expressions through elevated perception of active
listening. To test this possibility, we ran mediation analyses using PROCESS
macro (Model 4) \cite{hayes2013introduction} on (a) change in emotional state,
(b) perceived emotional support and (c) self-disclosure behaviors.  Taking into
consideration the pre- and post-measurement structure of (a) emotional states,
we first ran some preliminary repeated-measures analyses.  Keeping in mind the
significant effect of backchanneling on perceived active listening (IV to
mediator), but no significant effects of backchanneling on any of the emotional
changes (IV to DV), we first checked if emotional states changed based on
changes in active listening perceptions (mediator to DV).

We found that pre to post changes were significantly affected by
active listening perceptions  only for enthusiasm (p = .047), but not for the other 9 emotions
(ps $>$ .05). When a mediation analysis was run for enthusiasm as the DV
(calculated by deducting pre- from post-enthusiasm states), significant indirect effects emerged (indirect effect = .42, 95\% bias-corrected
10,000 bootstrap CI [0.0046, 1.0053]; see Figure \ref{fig:result-enthusiasm}),
 suggesting that \textbf{backchanneling increased the feeling of
enthusiasm due to heightened perception of active listening}.

For (b) perceived emotional support, the mediation analysis also suggested that
\textbf{backchanneling (vs. control) led to increased emotional support from
Alexa, by increasing user perceptions of active listening} (indirect effect
= .60, 95\% biased-corrected 10,000 bootstrap CI [0.0297, 1.4566]; see Figure
\ref{fig:result-perceived-support}). Similarly, for (c) self-disclosure
behaviors, \textbf{active listening perception mediated the positive
backchanneling effects on increased use of positive emotional words} (indirect
effects = .29, 95\% biased- corrected 10,000 bootstrap CI [0.0070, 0.8539]; see
Figure \ref{fig:result-positive-words}). Together, these significant mediation
effects indicate that, while the effects of backchanneling on emotional support
and positive self-disclosure did not occur directly, the effects took place
indirectly through (mediated by) active listening perceptions. That is, when
participants considered smart speakers as active listeners due to
backchanneling, they also perceived these devices as more emotionally
supportive. Similarly, the perception of smart speakers as active listeners
was associated with increased use of positive words during self-disclosure.

However, the significant main effect of backchanneling on the use of negative
emotional words was not explained by enhanced active listening perception
(indirect effects = -.05, 95\% biased-corrected 10,000 bootstrap CI [-.2045,
.0491]; see Figure \ref{fig:result-negative-words}). \markedchange{In addition, active
listening perception} did not serve as a statistically significant mediator for
interaction duration (indirect effects = -.33, 95\% biased-corrected 10,000
bootstrap CI [-63.21, 44.19]) nor word count (indirect effects = 3.75, 95\%
biased-corrected 10,000 bootstrap CI [-144.19, 107.29]).

\subsection{Serial mediation toward enhanced usability and other supplementary usability effects}

Given the significant and positive mediation effects of backchanneling on
perceived emotional support and the use of positive emotional words, we
explored if those effects would result in improved usability of the smart
speaker service (RQ1). A serial mediation analysis, based on PROCESS Model 6
\cite{hayes2013introduction}, suggested that \textbf{increased perceived
emotional support is positively related to attitudes toward Alexa}
(indirect effect = .35, 95\% biased-corrected 10,000 bootstrap CI [0.0149,
0.9645]; see Figure \ref{fig:result-support-usability}). 
However, the expression of more positive emotional words did not enhance perceived usability of Alexa (indirect effect = .02, 95\% biased-corrected 10,000 bootstrap CI[-0.1693, 0.3737]; see Figure \ref{fig:result-words-usability}).

When we analyzed the supplementary usability measures, we found that
participants felt interrupted by the use of pseudo-random backchanneling.
Specifically, participants in the control group (M = 2.25, SD = 1.71) scored
lower on irritation compared to the active listening group (M = 3.70, SD =
2.45) (t(38) = 2.17, p = .04).  This finding suggests that allocating space for
silence could also help to prevent interruptions and consequent irritation.
Even though research on social robots has been progressing in this domain
\cite{lala2017attentive}, smart speakers have not deployed these features.
However, user perceptions of other aspects of Alexa's responses (e.g.,
interaction length, appropriateness) were not significantly influenced by
backchanneling (ps $>$ .48).

In summary, interacting with Alexa decreased some negative emotions such as
sadness, nervousness, and anxiety in both control and active listening
conditions. More importantly, adding verbal backchanneling
caused participants to perceive Alexa as a much more active listener, \markedchange{which in turn, positively affected
their perceived emotional support and expression of positive emotions. In particular, when
users felt like they received emotional support through an actively
listening Alexa, it led to an improved user experience}.

\subsection{Qualitative findings}

We also conducted interviews at the end of the study. Given the exploratory nature of the study, we decided to use semi-structured interviews. During the interview session, we focused on understanding  participants' experiences with Alexa, their expectations regarding our skill, and usefulness of the skill in the context of therapy and counseling (see the supplementary document for the semi-structured interview guide).  Furthermore, we asked their ideas about how smart speakers can potentially be used to support mental health and wellbeing. These interviews were audio-recorded.  Following each interview, one author compiled transcripts of a participant’s responses.

To analyze interview data, one author performed a bottom-up thematic analysis to identify common themes in the dataset using a qualitative interpretivist approach described by Braun and Clarke \cite{ braun2012thematic}. A second author checked the generated themes against data to ensure consistency and avoid any potential biases. \markedchange{In case of disagreement, researchers engaged in a follow-up discussion to reach consensus}. In the following section, we describe the resulting final themes, which converged into 3 major ones. The first theme was related to the efficacy of using smart speakers for self-disclosure (N = 5). The second theme discussed the potential of utilizing backchanneling cues as effective verbal continuers, including the positive evaluation of the pseudo-random backchanneling cues (N = 3), and the need for backchanneling for those in the control condition (N = 5). Negative user inputs emerged as a third theme. For some users, the pseudo-random timing was was disruptive (N = 2). Others expressed needs for a more tailored and empathetic response from the smart speaker (N = 5). Future directions and potential use suggested by users are also added as part of our qualitative findings (N = 4). We have labeled participants in the active listening condition as P1--P20 and those in the control condition as P21--P40.

\subsubsection{Smart speakers can effectively support self-disclosure}

A key barrier to self-disclosure is the fear of negative evaluation and
judgement. A number of participants noted that a smart speaker can help address
this barrier given that it is a device that does not have the capacity to judge
them. For example, P16 commented that ``\emph{Alexa is not a person, it's
anonymous, it's like you can talk to it more. You feel more comfortable}.''
Similarly, P10 noted that ``\emph{the good thing is that there is no human
around so that you can reveal whatever there is about your personal stuff}.''
This perceived ease of disclosure to smart speaker is consistent with prior
studies involving virtual agents \cite{lucas2014s}.

Participants also commented about the positive outcomes of disclosure to smart
speakers. P15 commented about her interaction with smart speakers during the
study: ``\emph{\markedchange{what I liked [that it]} made me think about different
things, which is interesting that [a device] can make you think of something
that you wouldn't think of otherwise}.'' P14 thought her interaction with our
system was similar to talking to her real-life therapist: ``\emph{She [Alexa] is
so close to my first therapist [\ldots] it was good to talk. The less the
therapist talks the more like a mirror. You see your own reactions}''.

\subsubsection{Pseudo-random backchanneling as effective verbal continuers}
Participants in the control condition --- without any backchanneling ---
repeatedly pointed out the need for having verbal continuers. P32 from the
control condition suggested to ``\emph{add some cues between when I'm speaking
[\ldots] Like if I say something emotional you would say `uhum' or something
like that. Or that `I understand'. Some dynamic \markedchange{responses [are necessary],} instead of just
listening}''. P30 also commented about the lack of perceived \markedchange{active
listening}: ``\emph{there could be things like `yeah?', `aha', `I am listening'
to show that she is listening}''. These findings clearly show the need for
backchanneling to support self-disclosure.

For most participants in the active listening condition, the pseudo-random
backchanneling was effective as verbal continuers. P3 commented that ``\emph{it
responded to what I was saying ---`aha' like it was a real person}''.  As
intended, these backchanneling cues helped to promote the perception that smart
speakers are active listeners: ``\emph{[backchanneling] gives you the cue
that Alexa was listening to me. Like uhum, yes, \ldots}'' (P10). Another
participant noted the lack of backchanneling in current devices and preferred
our system: ``\emph{Yeah it is really interesting that you feel it is listening
to you. [With Siri] you have to say your statement completely and then you pause
for the reaction. And then you speak whatever you wanna say. But with this one,
it is more continuous. You don't need to stop and start and stop. I think it's
really good you have something [like this]}'' (P20).

However, pseudo-random
backchanneling can also inadvertently interrupt users.  P07 noted that
``\emph{[The other backchanneling cues] fit to what I was saying but this one
`hm' stood to me as weird. It was kinda like [\ldots] out of time}''. The
backchanneling cues out of order can also be problematic: ``\emph{[Alexa] kinda
interjected by saying `okay' and it was kinda jarring because I thought I had
done something to trigger her}'' (P18). To address these issues, future work
should develop other heuristics beyond just pre-determined time interval for
backchanneling cues. For example, long duration of silence or pause from the
user can potentially indicate an opportunity to deliver backchanneling cues and
verbal continuers. For this, it will be important to identify optimal pause
intervals in user speech for backchanneling cues.

\subsubsection{Need for more elaborate and empathetic responses} Participants
also expressed a need for more elaborate and empathetic responses from smart
speakers beyond simple backchanneling cues used in the study. P10 suggested
longer responses: ``\emph{I wish that [the responses] even longer than that
[\ldots] longer statements could motivate me to talk more and to share.}''
Participants also wanted smart speakers to adapt to the content and
characteristics of their self-disclosure.  P07 expected interactions to reflect
prior disclosure: ``\emph{if it had asked about specific points that I had said
[..] then it would have definitely felt like it was more involved.}''

Similarly, participants also thought more empathetic responses from smart
speakers will be useful in this context. P18 suggested ``\emph{saying phrases
like, `oh I understand', `oh that must be tough' \ldots, [being] more
empathetic}.'' P33 from the control condition commented that ``\emph{when I say, `this has impacted me
emotionally' [\ldots] I expected [responses] like `I feel you'}.''

However, delivering such personalized and context-specific responses will
require understanding the content and emotion disclosed by the user. This can be
challenging to achieve, particularly in a privacy preserving manner.

\subsubsection{Suggested future directions from participants} A number of
participants were very positive about the use of smart speakers to deliver
mental health interventions. P37 commented that ``\emph{even though I have a
therapist and a psychiatrist I can't see them every day. Of course, I only go
once every two weeks or something. You could just have Alexa at home to talk to
her whenever}''. P14 pointed out that immigrants can have significant
difficulties in reaching out to a therapist in a different country due to
language barrier and suggested that smart speakers can be used to address this:
``\emph{I was thinking one of the biggest problems in therapy is that if you are
living in another country you can't always find the therapist you can talk to.
Let's say you see your therapist [and they] decide what therapy you need, like
maybe you need more cognitive or behaviorist, and they're going to prescribe it
and then Alexa will take it from there. She can practice it with you in whatever
language you want. [\ldots] I've been thinking about it for a couple of years
actually. Do we really need to be in the same room as a therapist?}''.

Participants also suggested that social isolation and loneliness can potentially
be addressed by smart speakers. P10 commented that: ``\emph{It could be very
useful especially for those who spend most of their time alone and need
companions. Even for those who say they don't need a companion; I argue that it
will be still interesting for them to try such an agent. Because it is not just
about talking to another human being. Even imagining that something is listening
to you would actively engage your mind and [help to] reflect on your own.}'' P15
also suggested that our system based on smart speakers can benefit the
older population to address their social isolation \cite{nicholson2012review}.

Overall, there was a persistent theme regarding the need for an \markedchange{active
listener} during the course of one's daily activities. As P18 commented: ``\emph{sometimes we
can't get her [their daughter] to sleep and you just wanna like\ldots scream
into a pillow or something right? [\ldots] to get it off your chest [\ldots]
Being able to express that and have some level of interaction back is useful.
[\ldots] Whether it's for companionship, or meditative, or therapeutic, to feel
like someone is listening}''.

\section{Discussion} In this study, we implemented and empirically evaluated the
use of smart speakers as active listeners for supporting disclosure. More
specifically, we developed Alexa skills with pseudo-random backchanneling to
indicate active listening behaviors. Our findings show that backchanneling
can improve the degree of perceived active listening by smart speakers.
Furthermore, perceived active listening by smart speakers can foster
self-disclosure, provide emotional support, and improve usability. In this
section, we will discuss the implications of these findings as well as identify research opportunities for the CSCW and the broader HCI community.

\subsection{Implications for design and usability of smart speakers}
Our findings offer important design implications which can help to advance recent research trends in the HCI community regarding smart speaker interaction design \cite{10.1145/3411810,10.1145/3290605.3300705}. Specifically, these findings can extend interaction models currently supported by
smart speakers.  Users' engagement with smart speakers are now mostly limited to
task-specific commands with the resultant dialogues remaining short and often not spanning
more than a single turn. A smart speaker with active listening can enable
different types of interactions with users wherein dialogues are more open-ended,
longer in duration, and span multiple turns, which can be
particularly useful in sustaining user engagement. For instance, multi-turn dialogues can
improve ``message interactivity'' --- the degree of
interconnected and threaded conversations, which led to better engagement and
positive persuasive outcomes \cite{bellur2017talking}.
In addition, active listening can promote a smoother and more efficient acclimation process in broadening the use of smart speakers among certain age groups. For instance, older adults with less technological experience are hesitant to use smart speakers because of reliability concerns (e.g., reluctance to rely on Alexa for important reminders) \cite{10.1145/3373759}, as opposed to children, who are more prone to personify and emotionally associate with smart speakers \cite{10.1145/3196709.3196772,10.1145/3381002}. Thus, future work should explore other interactive and conversational features beyond basic continuers that can help to
transform smart speakers from mere task agents to companions with perceived
empathy across different age groups (e.g., \cite{qiu2021nurse}).

Perception of active listening in smart speakers can also help to improve
usability. \markedchange{Not only did our study empirically demonstrated that active listening perception served as a precursor to enhanced usability, participants in the control study condition also repeatedly reported that the lack of backchanneling cues to be a serious usability issue}. In particular, the positive effects of active listening on
usability were led by increased emotional support and the use of positive (but
not negative) emotional words. This shows how even simple interface signals of
active listening, such as backchanneling cues employed in this study, can
generate positive thoughts as well as improve usability. This seems to be a significant cue effect considering the primarily transactional nature of conversations between users and conversational agents, and the reluctance many users express in terms of building a social relationship with these agents \cite{10.1145/3290605.3300705}. These findings are consistent with Clark et al. \cite{10.1145/3290605.3300705}, which also noted  the importance of ``the role of the agent as a listener''.

That said, our findings indicate that the design of smart speakers with
active listening must be consistent with user expectations, limit
interruptions, and enable a supportive environment for continuing dialogue. Luger and Sellen \cite{10.1145/2858036.2858288} pointed out that especially for users with less technical knowledge, the large gap between expectations and experience led users to become less tolerant of system failures and limited \markedchange{their use of} smart speaker use. In our study, some participants noted that mistimed and unexpected backchanneling can
interrupt user focus and interaction. Future studies should work on developing
heuristics for backchanneling that can be used to avoid such user
interruptions.  For example, a long pause from a user can potentially indicate
an opportunity to deliver backchanneling cues encouraging them to continue
their interaction. This will require identifying appropriate pause intervals in
user speech to deliver backchanneling cues. In addition, changes in tone and
other non-content qualities of user interaction with the smart speaker could be
explored as potential triggers for introducing backchanneling cues into the
conversation. Furthermore, design efforts can be geared toward conveying
backchanneling via non-auditory means (\markedchange{e.g., using light behaviors} \cite{kunchay2019investigating,kunchay2021assessing}) which can be not only be more subtle and less intrusive, but also more
accessible, especially for hearing-impaired users
\cite{blair2019understanding,blair2020didn}.

Our study participants also noted the need for context- and sentiment-specific
backchanneling cues and verbal continuers \markedchange{(e.g., a smart speaker replying ``I am sorry to hear that’’ after a user discloses a negative experience)}.
There exist many psychological, physical, and social contextual factors that deter users from actively engaging with smart speakers, which calls for context-sensitive conversation management features \cite{10.1145/3411810}. Studies with chatbots have shown that such expressions of sympathy, even coming
from a machine, is perceived as more supportive by users facing a health issue
than straightforward provision of information and advice \cite{liu2018should}.
However, this can be challenging since it will involve real-time understanding
of contexts and appropriate classification of sentiments conveyed by the users.
Given the limited computational capabilities of smart speakers, this might
require offloading data processing to servers, which will create considerable
time lag as well as raise serious privacy concerns. To address these issues,
future work should focus on more optimized data processing in the devices to
identify contexts and sentiment. This will involve both speech recognition
(e.g., detecting keywords indicating sentiment and contexts) as well as audio
analysis (e.g., pitch indicating positive sentiment).

\subsection{Applications to therapy and mental health support} 
Given the therapeutic benefit of disclosure, our findings also have important implications for mental health interventions. Smart speakers with active listening can help
to engage users in regular self-disclosure practices, and 
such disclosure can lead to positive wellbeing outcomes as prior studies show \markedchange{\cite{frattaroli2006experimental,pennebaker1997writing,baikie2012expressive,krpan2013everyday,bugg2009randomised,craft2013expressive}}. From our qualitative interviews, we found that some of our participants were particularly positive about the potential use of smart
speakers to remotely deliver therapeutic interventions. They suggested the use of
smart speakers to provide just-in-time interventions and supplement their
(traditionally infrequent) interactions with mental health experts.

In addition, smart speaker technology can help to address emotional fatigue and burnout faced by
therapists. Vicarious and secondary trauma can lead
to emotional fatigue and burnout among therapists and mental health
practitioners \cite{newell2010professional}. Their job often requires exposing themselves to clients' most
difficult emotions, which can significantly impact their own emotional and
mental wellbeing \cite{rauvola_compassion_2019,newell2010professional}. As our
qualitative findings indicate, users are willing to use smart speakers to share
intimate feelings, especially when in-person therapy is unavailable at the
moment. This suggests that therapists can potentially use smart speakers as
assistive tools to continue facilitating effective disclosure from patients.
By integrating smart speakers in their workflow, therapists
thus can potentially reduce their continuous exposure to traumatic disclosure.

However, there are still considerable challenges that need to be addressed
before smart speakers can be used to successfully deliver mental health
interventions. Interactions supported by smart speakers are
considerably different from current eHealth practices using webpages and mobile
apps. As such, existing messages and interactions will need
to be adapted for smart speakers. For instance,
participants in our study called for more adaptive and empathetic responses based
on users' specific input. Furthermore, sustaining engagement is often
critical for successful mental health interventions. However, not much is known
about how to design smart speaker interactions to keep users engaged over a
period of time.  It will be critical to design and compare different
interaction patterns in smart speakers to identify their strengths and trade-offs
relevant to sustaining engagement.


\subsection{Trust, privacy and ethical concerns}
While smart speakers are becoming increasingly popular, there is a lack of
transparency regarding how these devices collect, handle, and process user
data. Given that disclosure can contain intimate and personal information, it
is essential to understand the risks of self-disclosure to commercialized smart
speakers. As such, users' trust and privacy concerns must be addressed before
smart speakers can be used to support disclosure and enable
therapeutic interventions. Our approach aimed to address such privacy issues by using ``context free''
interactions --- not using or recording any user content to deliver
backchanneling cues. Recently, Amazon
introduced Health Insurance Portability and Accountability Act (HIPAA) compliant
Alexa skills \cite{farr2019cnbc}. Similar steps can be used to ensure users' privacy regarding their disclosed information to an active smart speaker. In this regard, this study offers implications regarding protecting users' and patients' safety when utilizing conversational user interfaces (CUIs) for therapeutic self-disclosure via active listening (mostly studied in the chatbot context), which has been an ongoing challenge in fostering human-to-CUI collaboration within the CSCW community  \cite{10.1145/3406865.3418587}. Beyond supporting dyadic interactions between CUIs and users, this study also suggests expanded collaboration between those dyads and professional or informal health care providers (e.g., therapists, caretakers), by sharing the responsibilities and supporting collaborative tasks. 

A system focusing on therapeutic support must address ethical and user safety concerns as well \cite{han2021ptsdialogue}. For instance, users may feel deceived after finding out the limitations of smart speaker capabilities to provide support. \markedchange{Weizenbaum, the creator of one of the earliest chatbots (Eliza), noted the danger in ``how easy it is to create and maintain the illusion of understanding" (p. 42) \cite{Eliza}}. Similar to Eliza, our skill concealed its lack of understanding, which can create technical issues beyond ethical concerns. As Weizenbaum stated, while "it has a crucial psychological utility in that it serves the speaker to maintain his sense of being heard and understood" (p. 42), "to encourage its conversational partner to offer inputs from which it can select remedial information, it must reveal its misunderstanding" (p. 43) \cite{Eliza}. Misunderstanding itself can evoke user safety concerns as well. For instancce, without being able to detect users' disclosure to engage in harmful activities, the system cannot offer urgent preventive support. Thus, going forward, it will be essential to identify use cases where active listening would not be appropriate, and merit deeper therapeutic interventions that are personalized and adapted to users' needs. Admittedly, there exist challenges to developing a context-aware and personalized system, while still preserving privacy. Future work can explore ways to detect users' emotional status (e.g., via specific keywords and user prompts) without compromising data security and offer richer active listening cues beyond backchanneling. In fact, Ward \cite{Ward1996UsingPC} found that vocal attributes such as low pitch can be sufficient to determine when the backchanneling should be offered while building a responsive dialog system. This will be a fruitful venue to explore considering the significant emotional effects we found from a single-session of backchanneling cues that were delivered devoid of any context awareness. In addition, a system aiming to support self-disclosure must explicitly convey its abilities and limitations. It should also ensure access to external resources (e.g., hotline numbers) similar to many apps (e.g., PTSD Coach \cite{pub.1073661085}) and conversational agents (e.g., Woebot \cite{fitzpatrick2017delivering}) focusing on mental health that face similar concerns of perceived ability in helping users in distress \cite{han2021ptsdialogue}.

\subsection{Study limitations and future directions}
In our study, we aimed to evaluate the effectiveness of pseudo-random
backchanneling to support effective self-disclosure. To do so, we used pre- and
post-questionnaires to measure subjects' emotional states before and after
their single-session interactions with Alexa. While this approach is consistent with prior work on examining the psychological and emotional effects of self-disclosure (e.g., \cite{ho2018psychological}), we acknowledge that extended disclosure sessions over
several weeks may be needed in order to observe sustained emotional changes. The single-session study design may have also led to only finding few significant effects, with seemingly small effects. For instance, despite the statistically significant difference between conditions, active listening perception was very low across conditions (3.23 out of 10-point scale), with only 1.57 point difference between conditions. In addition, while the use of negative words was directly enhanced by backchanneling cues (increased to 0.7 from 0.4 of words), the negative words per interaction only amounts up to about 4.7 words (vs. 2.7 words). However, the fact that we found any significant effects from a single-session study is highly encouraging. Prolonged and sustained interactions with smart speakers that can reach a wide user base can lead to more pronounced effects, which can be examined by future studies. Moreover, considering the massive diffusion of smart speakers and the millions of exchanges that users all over the world have with them on a daily basis, the modest effect noticed in our study \markedchange{can translate to hundreds of thousands of users perceiving greater emotional support and engaging in beneficial self-disclosure with their smart speakers}.

There are of course ethical concerns with the use of pseudo (non-receptive) smart speaker responses, as it constitutes a form of deception to users. Yet, we would like to note that the participants in our study had a reasonable understanding of the abilities of these devices in that they not only correctly identified that they were interacting with a device but also noted the ease of disclosure to a non-human entity, which is consistent with prior work \cite{lucas2014s}. Another related concern regarding the pseudo-random nature of the smart speakers’ empathetic response is that  no claim can be made to support its efficacy for therapeutic interventions, since a) the interactions did not specifically focus on talking about negative or traumatic life events, \markedchange{ and b) the responses were not} specifically catered to individuals’ therapeutic needs. Still, self-disclosure is an established intervention process for both clinical and non-clinical populations and events (e.g., \cite{frattaroli2006experimental}), and the importance of active listening for clinical populations has been well-established \cite{danby2009listeners,hutchbyActiveListeningFormulations2005}. Thus, there is room for future research to focus on specific clinical or therapeutic contexts and interactions in order to examine the potential of smart speakers for interventions in a more focused manner. While there may be other platforms  that can facilitate a more tailored and time-sensitive backchanneling to better promote self-disclosure, studying backchanneling effects in the context of smart speakers is meaningful in that \markedchange{our findings can be applicable to many other consumer devices supporting voice interactions. Specifically, there are ``hundreds of millions of Alexa-enabled devices'' \cite{amazoncom_inc_alexa_2020} and Google Assistant is now available on more than one billion devices \cite{huffman_heres_2019}. Our findings regarding backchanneling and self-disclosure are potentially relevant 
for this wide range of consumer devices}. As such, future studies should continue examining the therapeutic use of consumer products such as smart speakers which can offer broader application reach, compared to custom-developed systems.

\section{Conclusion}
This study shows how self-disclosure through smart speakers can have significant positive effects on health and wellbeing by empirically evaluating the use of smart speakers as active listeners. Specifically, we leveraged the Amazon Alexa framework to deliver pseudo-random backchanneling cues during user interactions to indicate active listening. Our findings show that backchanneling significantly improved the perceived level of active listening in smart speakers. It also resulted in more emotional disclosure with increased use of positive words from participants. Furthermore, perception of smart speakers as active listeners was positively associated with perceived emotional support, which in turn was significantly associated with participants’ positive attitudes toward smart speakers.

These findings can help us design more engaging and more usable smart speakers. For instance, we learned that the inclusion of simple interactive cues in the form of backchanneling can allow users to consider smart speakers as confidants to share emotions and also obtain emotional support. Such findings not only point us to the specific design elements we should incorporate in technology-based health interventions but also suggest opportunity areas (e.g., adaptive or personalized backchanneling; more complex interpersonal conversational cues to increase active listening perceptions). Our study also shows the feasibility of using smart speakers to provide therapeutic support. Given their wide adoption, smart speakers can play a crucial role in addressing mental health and wellbeing issues at scale. More generally, it appears that verbal backchanneling by smart speakers serves as a subtle, but powerful, social lubricant that can improve user experience regardless of interaction context.

\bibliographystyle{ACM-Reference-Format}
\bibliography{bibliography}


\begin{thebibliography}{80}


\ifx \showCODEN    \undefined \def \showCODEN     #1{\unskip}     \fi
\ifx \showDOI      \undefined \def \showDOI       #1{#1}\fi
\ifx \showISBNx    \undefined \def \showISBNx     #1{\unskip}     \fi
\ifx \showISBNxiii \undefined \def \showISBNxiii  #1{\unskip}     \fi
\ifx \showISSN     \undefined \def \showISSN      #1{\unskip}     \fi
\ifx \showLCCN     \undefined \def \showLCCN      #1{\unskip}     \fi
\ifx \shownote     \undefined \def \shownote      #1{#1}          \fi
\ifx \showarticletitle \undefined \def \showarticletitle #1{#1}   \fi
\ifx \showURL      \undefined \def \showURL       {\relax}        \fi
\providecommand\bibfield[2]{#2}
\providecommand\bibinfo[2]{#2}
\providecommand\natexlab[1]{#1}
\providecommand\showeprint[2][]{arXiv:#2}

\bibitem[\protect\citeauthoryear{Alcántara}{Alcántara}{2021}]%
        {alcantaraSmart2021}
\bibfield{author}{\bibinfo{person}{Ann-Marie Alcántara}.}
  \bibinfo{year}{2021}\natexlab{}.
\newblock \showarticletitle{Smart {{Speakers Go Beyond Waiting}} to {{Be
  Asked}}}.
\newblock  (\bibinfo{year}{2021}).
\newblock
\showISSN{0099-9660}
\urldef\tempurl%
\url{https://www.wsj.com/articles/smart-speakers-go-beyond-waiting-to-be-asked-11620640802}
\showURL{%
\tempurl}


\bibitem[\protect\citeauthoryear{{Amazon.com, Inc.}}{{Amazon.com,
  Inc.}}{[n.d.]}]%
        {amazoncom_inc_speech_nodate}
\bibfield{author}{\bibinfo{person}{{Amazon.com, Inc.}}}
  \bibinfo{year}{[n.d.]}\natexlab{}.
\newblock \bibinfo{title}{Speech {Synthesis} {Markup} {Language} ({SSML})
  {Reference} {\textbar} {Alexa} {Skills} {Kit}}.
\newblock
\newblock
\urldef\tempurl%
\url{https://developer.amazon.com/en-US/docs/alexa/custom-skills/speech-synthesis-markup-language-ssml-reference.html}
\showURL{%
\tempurl}
\newblock
\shownote{Library Catalog: developer.amazon.com.}


\bibitem[\protect\citeauthoryear{{Amazon.com, Inc.}}{{Amazon.com,
  Inc.}}{2020}]%
        {amazoncom_inc_alexa_2020}
\bibfield{author}{\bibinfo{person}{{Amazon.com, Inc.}}}
  \bibinfo{year}{2020}\natexlab{}.
\newblock \bibinfo{title}{"{Alexa}, what’s happening this week at {CES}?"}.
\newblock
\newblock
\urldef\tempurl%
\url{https://blog.aboutamazon.com/devices/alexa-whats-happening-this-week-at-ces}
\showURL{%
\tempurl}


\bibitem[\protect\citeauthoryear{Baikie, Geerligs, and Wilhelm}{Baikie
  et~al\mbox{.}}{2012}]%
        {baikie2012expressive}
\bibfield{author}{\bibinfo{person}{Karen~A Baikie}, \bibinfo{person}{Liesbeth
  Geerligs}, {and} \bibinfo{person}{Kay Wilhelm}.}
  \bibinfo{year}{2012}\natexlab{}.
\newblock \showarticletitle{Expressive writing and positive writing for
  participants with mood disorders: An online randomized controlled trial}.
\newblock \bibinfo{journal}{\emph{Journal of affective disorders}}
  \bibinfo{volume}{136}, \bibinfo{number}{3} (\bibinfo{year}{2012}),
  \bibinfo{pages}{310--319}.
\newblock
\newblock
\shownote{\url{https://doi.org/10.1016/j.jad.2011.11.032}.}


\bibitem[\protect\citeauthoryear{Balon and Rim{\'e}}{Balon and
  Rim{\'e}}{2016}]%
        {balon2016lexical}
\bibfield{author}{\bibinfo{person}{S{\'e}verine Balon} {and}
  \bibinfo{person}{Bernard Rim{\'e}}.} \bibinfo{year}{2016}\natexlab{}.
\newblock \showarticletitle{Lexical profile of emotional disclosure in socially
  shared versus written narratives}.
\newblock \bibinfo{journal}{\emph{Journal of Language and Social Psychology}}
  \bibinfo{volume}{35}, \bibinfo{number}{4} (\bibinfo{year}{2016}),
  \bibinfo{pages}{345--373}.
\newblock
\newblock
\shownote{\url{https://doi.org/10.1177/0261927X15603425}.}


\bibitem[\protect\citeauthoryear{Bavelas, Coates, and Johnson}{Bavelas
  et~al\mbox{.}}{2000}]%
        {conarrators}
\bibfield{author}{\bibinfo{person}{Janet~B Bavelas}, \bibinfo{person}{Linda
  Coates}, {and} \bibinfo{person}{Trudy Johnson}.}
  \bibinfo{year}{2000}\natexlab{}.
\newblock \showarticletitle{Listeners as co-narrators}.
\newblock \bibinfo{journal}{\emph{Journal of personality and social
  psychology}}  \bibinfo{volume}{79} (\bibinfo{year}{2000}),
  \bibinfo{pages}{941--952}.
\newblock
\urldef\tempurl%
\url{https://doi.org/"10.1037//0022-3514.79.6.941"}
\showDOI{\tempurl}


\bibitem[\protect\citeauthoryear{Bellur and Sundar}{Bellur and Sundar}{2017}]%
        {bellur2017talking}
\bibfield{author}{\bibinfo{person}{Saraswathi Bellur} {and}
  \bibinfo{person}{S.~Shyam Sundar}.} \bibinfo{year}{2017}\natexlab{}.
\newblock \showarticletitle{Talking health with a machine: How does message
  interactivity affect attitudes and cognitions?}
\newblock \bibinfo{journal}{\emph{Human Communication Research}}
  \bibinfo{volume}{43}, \bibinfo{number}{1} (\bibinfo{year}{2017}),
  \bibinfo{pages}{25--53}.
\newblock
\newblock
\shownote{\url{https://doi.org/10.1111/hcre.12094}.}


\bibitem[\protect\citeauthoryear{Blair and Abdullah}{Blair and
  Abdullah}{2019}]%
        {blair2019understanding}
\bibfield{author}{\bibinfo{person}{Johnna Blair} {and} \bibinfo{person}{Saeed
  Abdullah}.} \bibinfo{year}{2019}\natexlab{}.
\newblock \showarticletitle{Understanding the Needs and Challenges of Using
  Conversational Agents for Deaf Older Adults}. In
  \bibinfo{booktitle}{\emph{Conference Companion Publication of the 2019 on
  Computer Supported Cooperative Work and Social Computing}}.
  \bibinfo{pages}{161--165}.
\newblock


\bibitem[\protect\citeauthoryear{Blair and Abdullah}{Blair and
  Abdullah}{2020}]%
        {blair2020didn}
\bibfield{author}{\bibinfo{person}{Johnna Blair} {and} \bibinfo{person}{Saeed
  Abdullah}.} \bibinfo{year}{2020}\natexlab{}.
\newblock \showarticletitle{It Didn't Sound Good with My Cochlear Implants:
  Understanding the Challenges of Using Smart Assistants for Deaf and Hard of
  Hearing Users}.
\newblock \bibinfo{journal}{\emph{Proceedings of the ACM on Interactive,
  Mobile, Wearable and Ubiquitous Technologies}} \bibinfo{volume}{4},
  \bibinfo{number}{4} (\bibinfo{year}{2020}), \bibinfo{pages}{1--27}.
\newblock


\bibitem[\protect\citeauthoryear{Bodie, Vickery, Cannava, and Jones}{Bodie
  et~al\mbox{.}}{2015}]%
        {bodie_role_2015}
\bibfield{author}{\bibinfo{person}{Graham~D. Bodie}, \bibinfo{person}{Andrea~J.
  Vickery}, \bibinfo{person}{Kaitlin Cannava}, {and}
  \bibinfo{person}{Susanne~M. Jones}.} \bibinfo{year}{2015}\natexlab{}.
\newblock \showarticletitle{The {Role} of ``{Active} {Listening}'' in
  {Informal} {Helping} {Conversations}: {Impact} on {Perceptions} of {Listener}
  {Helpfulness}, {Sensitivity}, and {Supportiveness} and {Discloser}
  {Emotional} {Improvement}}.
\newblock \bibinfo{journal}{\emph{Western Journal of Communication}}
  \bibinfo{volume}{79}, \bibinfo{number}{2} (\bibinfo{year}{2015}),
  \bibinfo{pages}{151--173}.
\newblock
\showISSN{1057-0314}
\urldef\tempurl%
\url{https://doi.org/10.1080/10570314.2014.943429}
\showDOI{\tempurl}


\bibitem[\protect\citeauthoryear{Braun and Clarke}{Braun and Clarke}{2012}]%
        {braun2012thematic}
\bibfield{author}{\bibinfo{person}{Virginia Braun} {and}
  \bibinfo{person}{Victoria Clarke}.} \bibinfo{year}{2012}\natexlab{}.
\newblock \showarticletitle{Thematic analysis.}
\newblock  (\bibinfo{year}{2012}).
\newblock


\bibitem[\protect\citeauthoryear{Bugg, Turpin, Mason, and Scholes}{Bugg
  et~al\mbox{.}}{2009}]%
        {bugg2009randomised}
\bibfield{author}{\bibinfo{person}{Alison Bugg}, \bibinfo{person}{Graham
  Turpin}, \bibinfo{person}{Suzanne Mason}, {and} \bibinfo{person}{Cathy
  Scholes}.} \bibinfo{year}{2009}\natexlab{}.
\newblock \showarticletitle{A randomised controlled trial of the effectiveness
  of writing as a self-help intervention for traumatic injury patients at risk
  of developing post-traumatic stress disorder}.
\newblock \bibinfo{journal}{\emph{Behaviour Research and Therapy}}
  \bibinfo{volume}{47}, \bibinfo{number}{1} (\bibinfo{year}{2009}),
  \bibinfo{pages}{6--12}.
\newblock
\newblock
\shownote{\url{https://doi.org/10.1016/j.brat.2008.10.006}.}


\bibitem[\protect\citeauthoryear{Cha, Kim, Park, Kang, Park, Lee, Lee, and
  Lee}{Cha et~al\mbox{.}}{2020}]%
        {10.1145/3411810}
\bibfield{author}{\bibinfo{person}{Narae Cha}, \bibinfo{person}{Auk Kim},
  \bibinfo{person}{Cheul~Young Park}, \bibinfo{person}{Soowon Kang},
  \bibinfo{person}{Mingyu Park}, \bibinfo{person}{Jae-Gil Lee},
  \bibinfo{person}{Sangsu Lee}, {and} \bibinfo{person}{Uichin Lee}.}
  \bibinfo{year}{2020}\natexlab{}.
\newblock \showarticletitle{Hello There! Is Now a Good Time to Talk? Opportune
  Moments for Proactive Interactions with Smart Speakers}.
\newblock \bibinfo{journal}{\emph{Proc. ACM Interact. Mob. Wearable Ubiquitous
  Technol.}} \bibinfo{volume}{4}, \bibinfo{number}{3}, Article
  \bibinfo{articleno}{74} (\bibinfo{date}{Sept.} \bibinfo{year}{2020}),
  \bibinfo{numpages}{28}~pages.
\newblock
\urldef\tempurl%
\url{https://doi.org/10.1145/3411810}
\showDOI{\tempurl}


\bibitem[\protect\citeauthoryear{Clark, Pantidi, Cooney, Doyle, Garaialde,
  Edwards, Spillane, Gilmartin, Murad, Munteanu, Wade, and Cowan}{Clark
  et~al\mbox{.}}{2019}]%
        {10.1145/3290605.3300705}
\bibfield{author}{\bibinfo{person}{Leigh Clark}, \bibinfo{person}{Nadia
  Pantidi}, \bibinfo{person}{Orla Cooney}, \bibinfo{person}{Philip Doyle},
  \bibinfo{person}{Diego Garaialde}, \bibinfo{person}{Justin Edwards},
  \bibinfo{person}{Brendan Spillane}, \bibinfo{person}{Emer Gilmartin},
  \bibinfo{person}{Christine Murad}, \bibinfo{person}{Cosmin Munteanu},
  \bibinfo{person}{Vincent Wade}, {and} \bibinfo{person}{Benjamin~R. Cowan}.}
  \bibinfo{year}{2019}\natexlab{}.
\newblock \showarticletitle{What Makes a Good Conversation? Challenges in
  Designing Truly Conversational Agents}. In
  \bibinfo{booktitle}{\emph{Proceedings of the 2019 CHI Conference on Human
  Factors in Computing Systems}} (Glasgow, Scotland Uk)
  \emph{(\bibinfo{series}{CHI '19})}. \bibinfo{publisher}{Association for
  Computing Machinery}, \bibinfo{address}{New York, NY, USA},
  \bibinfo{pages}{1–12}.
\newblock
\showISBNx{9781450359702}
\urldef\tempurl%
\url{https://doi.org/10.1145/3290605.3300705}
\showDOI{\tempurl}


\bibitem[\protect\citeauthoryear{Clark, Pierce, Finn, Hsu, Toosley, and
  Williams}{Clark et~al\mbox{.}}{1998}]%
        {clark1998impact}
\bibfield{author}{\bibinfo{person}{Ruth~Anne Clark}, \bibinfo{person}{Amy~J
  Pierce}, \bibinfo{person}{Kathleen Finn}, \bibinfo{person}{Karen Hsu},
  \bibinfo{person}{Adam Toosley}, {and} \bibinfo{person}{Lionel Williams}.}
  \bibinfo{year}{1998}\natexlab{}.
\newblock \showarticletitle{The impact of alternative approaches to comforting,
  closeness of relationship, and gender on multiple measures of effectiveness}.
\newblock \bibinfo{journal}{\emph{Communication Studies}} \bibinfo{volume}{49},
  \bibinfo{number}{3} (\bibinfo{year}{1998}), \bibinfo{pages}{224--239}.
\newblock
\newblock
\shownote{\url{https://doi.org/10.1080/10510979809368533}.}


\bibitem[\protect\citeauthoryear{Cohn, Chen, and Yu}{Cohn
  et~al\mbox{.}}{2019}]%
        {cohn-etal-2019-large}
\bibfield{author}{\bibinfo{person}{Michelle Cohn}, \bibinfo{person}{Chun-Yen
  Chen}, {and} \bibinfo{person}{Zhou Yu}.} \bibinfo{year}{2019}\natexlab{}.
\newblock \showarticletitle{A Large-Scale User Study of an {A}lexa {P}rize
  Chatbot: Effect of {TTS} Dynamism on Perceived Quality of Social Dialog}. In
  \bibinfo{booktitle}{\emph{Proceedings of the 20th Annual SIGdial Meeting on
  Discourse and Dialogue}}. \bibinfo{publisher}{Association for Computational
  Linguistics}, \bibinfo{address}{Stockholm, Sweden},
  \bibinfo{pages}{293--306}.
\newblock
\urldef\tempurl%
\url{https://doi.org/10.18653/v1/W19-5935}
\showDOI{\tempurl}


\bibitem[\protect\citeauthoryear{Craft, Davis, and Paulson}{Craft
  et~al\mbox{.}}{2013}]%
        {craft2013expressive}
\bibfield{author}{\bibinfo{person}{Melissa~A Craft}, \bibinfo{person}{Gail~C
  Davis}, {and} \bibinfo{person}{Ren{\'e}~M Paulson}.}
  \bibinfo{year}{2013}\natexlab{}.
\newblock \showarticletitle{Expressive writing in early breast cancer
  survivors}.
\newblock \bibinfo{journal}{\emph{Journal of Advanced Nursing}}
  \bibinfo{volume}{69}, \bibinfo{number}{2} (\bibinfo{year}{2013}),
  \bibinfo{pages}{305--315}.
\newblock
\newblock
\shownote{\url{https://doi.org/10.1111/j.1365-2648.2012.06008.x}.}


\bibitem[\protect\citeauthoryear{Danby, Butler, and Emmison}{Danby
  et~al\mbox{.}}{2009}]%
        {danby2009listeners}
\bibfield{author}{\bibinfo{person}{Susan~J Danby}, \bibinfo{person}{Carly
  Butler}, {and} \bibinfo{person}{Michael Emmison}.}
  \bibinfo{year}{2009}\natexlab{}.
\newblock \showarticletitle{When `listeners can't talk': Comparing active
  listening in opening sequences of telephone and online counselling}.
\newblock \bibinfo{journal}{\emph{Australian Journal of Communication}}
  \bibinfo{volume}{36}, \bibinfo{number}{3} (\bibinfo{year}{2009}),
  \bibinfo{pages}{91--113}.
\newblock


\bibitem[\protect\citeauthoryear{DeVault, Sagae, and Traum}{DeVault
  et~al\mbox{.}}{2011}]%
        {devault2011incremental}
\bibfield{author}{\bibinfo{person}{David DeVault}, \bibinfo{person}{Kenji
  Sagae}, {and} \bibinfo{person}{David Traum}.}
  \bibinfo{year}{2011}\natexlab{}.
\newblock \showarticletitle{Incremental interpretation and prediction of
  utterance meaning for interactive dialogue}.
\newblock \bibinfo{journal}{\emph{Dialogue \& Discourse}} \bibinfo{volume}{2},
  \bibinfo{number}{1} (\bibinfo{year}{2011}), \bibinfo{pages}{143--170}.
\newblock
\newblock
\shownote{\url{https://doi.org/10.5087/dad.2011.107}.}


\bibitem[\protect\citeauthoryear{Farr}{Farr}{2019}]%
        {farr2019cnbc}
\bibfield{author}{\bibinfo{person}{Christina Farr}.}
  \bibinfo{year}{2019}\natexlab{}.
\newblock \bibinfo{title}{`Alexa, find me a doctor': Amazon Alexa adds new
  medical skills}.
\newblock
\newblock
\newblock
\shownote{\url{https://www.cnbc.com/2019/04/03/amazon-alexa-hipaa-compliant-adds-medical-skills.html}.}


\bibitem[\protect\citeauthoryear{Fitzpatrick, Darcy, and Vierhile}{Fitzpatrick
  et~al\mbox{.}}{2017}]%
        {fitzpatrick2017delivering}
\bibfield{author}{\bibinfo{person}{Kathleen~Kara Fitzpatrick},
  \bibinfo{person}{Alison Darcy}, {and} \bibinfo{person}{Molly Vierhile}.}
  \bibinfo{year}{2017}\natexlab{}.
\newblock \showarticletitle{Delivering cognitive behavior therapy to young
  adults with symptoms of depression and anxiety using a fully automated
  conversational agent (Woebot): a randomized controlled trial}.
\newblock \bibinfo{journal}{\emph{JMIR mental health}} \bibinfo{volume}{4},
  \bibinfo{number}{2} (\bibinfo{year}{2017}), \bibinfo{pages}{e19}.
\newblock
\newblock
\shownote{\url{https://doi.org/10.2196/mental.7785}.}


\bibitem[\protect\citeauthoryear{Frattaroli}{Frattaroli}{2006}]%
        {frattaroli2006experimental}
\bibfield{author}{\bibinfo{person}{Joanne Frattaroli}.}
  \bibinfo{year}{2006}\natexlab{}.
\newblock \showarticletitle{Experimental disclosure and its moderators: a
  meta-analysis.}
\newblock \bibinfo{journal}{\emph{Psychological bulletin}}
  \bibinfo{volume}{132}, \bibinfo{number}{6} (\bibinfo{year}{2006}),
  \bibinfo{pages}{823}.
\newblock
\newblock
\shownote{\url{https://doi.org/10.1037/0033-2909.132.6.823}.}


\bibitem[\protect\citeauthoryear{Gardner}{Gardner}{2001}]%
        {gardner2001listeners}
\bibfield{author}{\bibinfo{person}{Rod Gardner}.}
  \bibinfo{year}{2001}\natexlab{}.
\newblock \showarticletitle{When listeners talk: Response tokens and listener
  stance.}
\newblock  (\bibinfo{year}{2001}).
\newblock
\newblock
\shownote{\url{https://doi.org/10.1075/pbns.92}.}


\bibitem[\protect\citeauthoryear{Garg and Sengupta}{Garg and Sengupta}{2020}]%
        {10.1145/3381002}
\bibfield{author}{\bibinfo{person}{Radhika Garg} {and}
  \bibinfo{person}{Subhasree Sengupta}.} \bibinfo{year}{2020}\natexlab{}.
\newblock \showarticletitle{He Is Just Like Me: A Study of the Long-Term Use of
  Smart Speakers by Parents and Children}.
\newblock \bibinfo{journal}{\emph{Proc. ACM Interact. Mob. Wearable Ubiquitous
  Technol.}} \bibinfo{volume}{4}, \bibinfo{number}{1}, Article
  \bibinfo{articleno}{11} (\bibinfo{date}{March} \bibinfo{year}{2020}),
  \bibinfo{numpages}{24}~pages.
\newblock
\urldef\tempurl%
\url{https://doi.org/10.1145/3381002}
\showDOI{\tempurl}


\bibitem[\protect\citeauthoryear{Gearhart and Bodie}{Gearhart and
  Bodie}{2011}]%
        {gearhart2011active}
\bibfield{author}{\bibinfo{person}{Christopher~C Gearhart} {and}
  \bibinfo{person}{Graham~D Bodie}.} \bibinfo{year}{2011}\natexlab{}.
\newblock \showarticletitle{Active-empathic listening as a general social
  skill: Evidence from bivariate and canonical correlations}.
\newblock \bibinfo{journal}{\emph{Communication Reports}} \bibinfo{volume}{24},
  \bibinfo{number}{2} (\bibinfo{year}{2011}), \bibinfo{pages}{86--98}.
\newblock
\newblock
\shownote{\url{https://doi.org/10.1080/08934215.2011.610731}.}


\bibitem[\protect\citeauthoryear{Han, Mendu, Jaworski, E~Owen, and
  Abdullah}{Han et~al\mbox{.}}{2021}]%
        {han2021ptsdialogue}
\bibfield{author}{\bibinfo{person}{Hee~Jeong Han}, \bibinfo{person}{Sanjana
  Mendu}, \bibinfo{person}{Beth~K Jaworski}, \bibinfo{person}{Jason E~Owen},
  {and} \bibinfo{person}{Saeed Abdullah}.} \bibinfo{year}{2021}\natexlab{}.
\newblock \bibinfo{booktitle}{\emph{PTSDialogue: Designing a Conversational
  Agent to Support Individuals with Post-Traumatic Stress Disorder}}.
\newblock \bibinfo{publisher}{Association for Computing Machinery},
  \bibinfo{address}{New York, NY, USA}, \bibinfo{pages}{198--203}.
\newblock
\showISBNx{9781450384612}
\urldef\tempurl%
\url{https://doi.org/10.1145/3460418.3479332}
\showURL{%
\tempurl}


\bibitem[\protect\citeauthoryear{Hayes}{Hayes}{2013}]%
        {hayes2013introduction}
\bibfield{author}{\bibinfo{person}{Andrew~F Hayes}.}
  \bibinfo{year}{2013}\natexlab{}.
\newblock \showarticletitle{Introduction to mediation, moderation, and
  conditional process analysis: A regression-based approach.}
\newblock  (\bibinfo{year}{2013}).
\newblock


\bibitem[\protect\citeauthoryear{Ho, Hancock, and Miner}{Ho
  et~al\mbox{.}}{2018}]%
        {ho2018psychological}
\bibfield{author}{\bibinfo{person}{Annabell Ho}, \bibinfo{person}{Jeff
  Hancock}, {and} \bibinfo{person}{Adam~S Miner}.}
  \bibinfo{year}{2018}\natexlab{}.
\newblock \showarticletitle{Psychological, relational, and emotional effects of
  self-disclosure after conversations with a chatbot}.
\newblock \bibinfo{journal}{\emph{Journal of Communication}}
  \bibinfo{volume}{68}, \bibinfo{number}{4} (\bibinfo{year}{2018}),
  \bibinfo{pages}{712--733}.
\newblock
\newblock
\shownote{\url{https://doi.org/10.1093/joc/jqy026}.}


\bibitem[\protect\citeauthoryear{Huffman}{Huffman}{2019}]%
        {huffman_heres_2019}
\bibfield{author}{\bibinfo{person}{Scott Huffman}.}
  \bibinfo{year}{2019}\natexlab{}.
\newblock \bibinfo{title}{Here’s how the {Google} {Assistant} became more
  helpful in 2018}.
\newblock
\newblock
\urldef\tempurl%
\url{https://blog.google/products/assistant/heres-how-google-assistant-became-more-helpful-2018/}
\showURL{%
\tempurl}


\bibitem[\protect\citeauthoryear{Hutchby}{Hutchby}{2005}]%
        {hutchbyActiveListeningFormulations2005}
\bibfield{author}{\bibinfo{person}{Ian Hutchby}.}
  \bibinfo{year}{2005}\natexlab{}.
\newblock \showarticletitle{"{{Active Listening}}": {{Formulations}} and the
  {{Elicitation}} of {{Feelings}}-{{Talk}} in {{Child Counselling}}}.
\newblock \bibinfo{journal}{\emph{Research on Language and Social Interaction}}
  \bibinfo{volume}{38}, \bibinfo{number}{3} (\bibinfo{date}{July}
  \bibinfo{year}{2005}), \bibinfo{pages}{303--329}.
\newblock
\showISSN{0835-1813}
\urldef\tempurl%
\url{https://doi.org/10.1207/s15327973rlsi3803_4}
\showDOI{\tempurl}


\bibitem[\protect\citeauthoryear{Jagosh, Donald~Boudreau, Steinert, MacDonald,
  and Ingram}{Jagosh et~al\mbox{.}}{2011}]%
        {jagosh_importance_2011}
\bibfield{author}{\bibinfo{person}{Justin Jagosh}, \bibinfo{person}{Joseph
  Donald~Boudreau}, \bibinfo{person}{Yvonne Steinert},
  \bibinfo{person}{Mary~Ellen MacDonald}, {and} \bibinfo{person}{Lois Ingram}.}
  \bibinfo{year}{2011}\natexlab{}.
\newblock \showarticletitle{The importance of physician listening from the
  patients’ perspective: {Enhancing} diagnosis, healing, and the
  doctor–patient relationship}.
\newblock \bibinfo{journal}{\emph{Patient Education and Counseling}}
  \bibinfo{volume}{85}, \bibinfo{number}{3} (\bibinfo{year}{2011}),
  \bibinfo{pages}{369--374}.
\newblock
\showISSN{0738-3991}
\urldef\tempurl%
\url{https://doi.org/10.1016/j.pec.2011.01.028}
\showDOI{\tempurl}


\bibitem[\protect\citeauthoryear{Jeong, Lee, and Kang}{Jeong
  et~al\mbox{.}}{2019}]%
        {jeong2019exploring}
\bibfield{author}{\bibinfo{person}{Yuin Jeong}, \bibinfo{person}{Juho Lee},
  {and} \bibinfo{person}{Younah Kang}.} \bibinfo{year}{2019}\natexlab{}.
\newblock \showarticletitle{Exploring Effects of Conversational Fillers on User
  Perception of Conversational Agents}. In \bibinfo{booktitle}{\emph{Extended
  Abstracts of the 2019 CHI Conference on Human Factors in Computing Systems}}.
  \bibinfo{pages}{1--6}.
\newblock
\newblock
\shownote{\url{https://doi.org/10.1145/3290607.3312913}.}


\bibitem[\protect\citeauthoryear{Johansson, Hori, Skantze, H{\"o}thker, and
  Gustafson}{Johansson et~al\mbox{.}}{2016}]%
        {johansson2016making}
\bibfield{author}{\bibinfo{person}{Martin Johansson}, \bibinfo{person}{Tatsuro
  Hori}, \bibinfo{person}{Gabriel Skantze}, \bibinfo{person}{Anja H{\"o}thker},
  {and} \bibinfo{person}{Joakim Gustafson}.} \bibinfo{year}{2016}\natexlab{}.
\newblock \showarticletitle{Making turn-taking decisions for an active
  listening robot for memory training}. In
  \bibinfo{booktitle}{\emph{International Conference on Social Robotics}}.
  Springer, \bibinfo{pages}{940--949}.
\newblock
\newblock
\shownote{\url{https://doi.org/10.1007/978-3-319-47437-3}.}


\bibitem[\protect\citeauthoryear{Jones}{Jones}{2004}]%
        {jones2004putting}
\bibfield{author}{\bibinfo{person}{Susanne Jones}.}
  \bibinfo{year}{2004}\natexlab{}.
\newblock \showarticletitle{Putting the person into person-centered and
  immediate emotional support: Emotional change and perceived helper competence
  as outcomes of comforting in helping situations}.
\newblock \bibinfo{journal}{\emph{Communication research}}
  \bibinfo{volume}{31}, \bibinfo{number}{3} (\bibinfo{year}{2004}),
  \bibinfo{pages}{338--360}.
\newblock
\newblock
\shownote{\url{https://doi.org/10.1177/0093650204263436}.}


\bibitem[\protect\citeauthoryear{Jones}{Jones}{2011}]%
        {jonesSupportiveListening2011}
\bibfield{author}{\bibinfo{person}{Susanne~M. Jones}.}
  \bibinfo{year}{2011}\natexlab{}.
\newblock \showarticletitle{Supportive {{Listening}}}.
\newblock \bibinfo{journal}{\emph{International Journal of Listening}}
  \bibinfo{volume}{25}, \bibinfo{number}{1-2} (\bibinfo{date}{Jan.}
  \bibinfo{year}{2011}), \bibinfo{pages}{85--103}.
\newblock
\showISSN{1090-4018, 1932-586X}
\urldef\tempurl%
\url{https://doi.org/10.1080/10904018.2011.536475}
\showDOI{\tempurl}


\bibitem[\protect\citeauthoryear{Kalyanaraman and Sundar}{Kalyanaraman and
  Sundar}{2006}]%
        {kalyanaraman2006psychological}
\bibfield{author}{\bibinfo{person}{Sriram Kalyanaraman} {and}
  \bibinfo{person}{S.~Shyam Sundar}.} \bibinfo{year}{2006}\natexlab{}.
\newblock \showarticletitle{The psychological appeal of personalized content in
  web portals: Does customization affect attitudes and behavior?}
\newblock \bibinfo{journal}{\emph{Journal of Communication}}
  \bibinfo{volume}{56}, \bibinfo{number}{1} (\bibinfo{year}{2006}),
  \bibinfo{pages}{110--132}.
\newblock
\newblock
\shownote{\url{https://doi.org/10.1111/j.1460-2466.2006.00006.x}.}


\bibitem[\protect\citeauthoryear{Kamita, Matsumoto, Sun, and Inoue}{Kamita
  et~al\mbox{.}}{2020}]%
        {kamita2020promotion}
\bibfield{author}{\bibinfo{person}{Takeshi Kamita}, \bibinfo{person}{Atsuko
  Matsumoto}, \bibinfo{person}{Boyu Sun}, {and} \bibinfo{person}{Tomoo Inoue}.}
  \bibinfo{year}{2020}\natexlab{}.
\newblock \showarticletitle{Promotion of Continuous Use of a Self-guided Mental
  Healthcare System by a Chatbot}. In \bibinfo{booktitle}{\emph{Conference
  Companion Publication of the 2020 on Computer Supported Cooperative Work and
  Social Computing}}. \bibinfo{pages}{293--298}.
\newblock
\urldef\tempurl%
\url{https://doi.org/10.1145/3406865.3418343}
\showURL{%
\tempurl}


\bibitem[\protect\citeauthoryear{Kawahara, Yamaguchi, Inoue, Takanashi, and
  Ward}{Kawahara et~al\mbox{.}}{2016}]%
        {kawahara2016prediction}
\bibfield{author}{\bibinfo{person}{Tatsuya Kawahara}, \bibinfo{person}{Takashi
  Yamaguchi}, \bibinfo{person}{Koji Inoue}, \bibinfo{person}{Katsuya
  Takanashi}, {and} \bibinfo{person}{Nigel~G Ward}.}
  \bibinfo{year}{2016}\natexlab{}.
\newblock \showarticletitle{Prediction and Generation of Backchannel Form for
  Attentive Listening Systems}. In \bibinfo{booktitle}{\emph{Interspeech}}.
  \bibinfo{pages}{2890--2894}.
\newblock
\newblock
\shownote{\url{https://doi.org/10.21437/Interspeech.2016-118}.}


\bibitem[\protect\citeauthoryear{Kornhaber, Walsh, Duff, and Walker}{Kornhaber
  et~al\mbox{.}}{2016}]%
        {kornhaber_enhancing_2016}
\bibfield{author}{\bibinfo{person}{Rachel Kornhaber}, \bibinfo{person}{Kenneth
  Walsh}, \bibinfo{person}{Jed Duff}, {and} \bibinfo{person}{Kim Walker}.}
  \bibinfo{year}{2016}\natexlab{}.
\newblock \showarticletitle{Enhancing adult therapeutic interpersonal
  relationships in the acute health care setting: an integrative review}.
\newblock \bibinfo{journal}{\emph{Journal of Multidisciplinary Healthcare}}
  \bibinfo{volume}{9} (\bibinfo{year}{2016}), \bibinfo{pages}{537--546}.
\newblock
\showISSN{1178-2390}
\urldef\tempurl%
\url{https://doi.org/10.2147/JMDH.S116957}
\showDOI{\tempurl}


\bibitem[\protect\citeauthoryear{Krpan, Kross, Berman, Deldin, Askren, and
  Jonides}{Krpan et~al\mbox{.}}{2013}]%
        {krpan2013everyday}
\bibfield{author}{\bibinfo{person}{Katherine~M Krpan}, \bibinfo{person}{Ethan
  Kross}, \bibinfo{person}{Marc~G Berman}, \bibinfo{person}{Patricia~J Deldin},
  \bibinfo{person}{Mary~K Askren}, {and} \bibinfo{person}{John Jonides}.}
  \bibinfo{year}{2013}\natexlab{}.
\newblock \showarticletitle{An everyday activity as a treatment for depression:
  the benefits of expressive writing for people diagnosed with major depressive
  disorder}.
\newblock \bibinfo{journal}{\emph{Journal of affective disorders}}
  \bibinfo{volume}{150}, \bibinfo{number}{3} (\bibinfo{year}{2013}),
  \bibinfo{pages}{1148--1151}.
\newblock
\newblock
\shownote{\url{https://doi.org/10.1016/j.jad.2013.05.065}.}


\bibitem[\protect\citeauthoryear{Kuhn, Greene, Hoffman, Nguyen, Wald, Schmidt,
  Ramsey, and Ruzek}{Kuhn et~al\mbox{.}}{2014}]%
        {pub.1073661085}
\bibfield{author}{\bibinfo{person}{Eric Kuhn}, \bibinfo{person}{Carolyn
  Greene}, \bibinfo{person}{Julia Hoffman}, \bibinfo{person}{Tam Nguyen},
  \bibinfo{person}{Laura Wald}, \bibinfo{person}{Janet Schmidt},
  \bibinfo{person}{Kelly~M Ramsey}, {and} \bibinfo{person}{Josef Ruzek}.}
  \bibinfo{year}{2014}\natexlab{}.
\newblock \showarticletitle{Preliminary evaluation of PTSD Coach, a smartphone
  app for post-traumatic stress symptoms.}
\newblock \bibinfo{journal}{\emph{Military Medicine}} \bibinfo{volume}{179},
  \bibinfo{number}{1} (\bibinfo{year}{2014}), \bibinfo{pages}{12--8}.
\newblock
\urldef\tempurl%
\url{https://doi.org/10.7205/milmed-d-13-00271}
\showDOI{\tempurl}


\bibitem[\protect\citeauthoryear{Kuhn, Bradbury, Nussbeck, and Bodenmann}{Kuhn
  et~al\mbox{.}}{2018}]%
        {kuhn_power_2018}
\bibfield{author}{\bibinfo{person}{Rebekka Kuhn}, \bibinfo{person}{Thomas~N.
  Bradbury}, \bibinfo{person}{Fridtjof~W. Nussbeck}, {and} \bibinfo{person}{Guy
  Bodenmann}.} \bibinfo{year}{2018}\natexlab{}.
\newblock \showarticletitle{The power of listening: {Lending} an ear to the
  partner during dyadic coping conversations}.
\newblock \bibinfo{journal}{\emph{Journal of Family Psychology}}
  \bibinfo{volume}{32}, \bibinfo{number}{6} (\bibinfo{year}{2018}),
  \bibinfo{pages}{762--772}.
\newblock
\showISSN{1939-1293, 0893-3200}
\urldef\tempurl%
\url{https://doi.org/10.1037/fam0000421}
\showDOI{\tempurl}


\bibitem[\protect\citeauthoryear{Kunchay and Abdullah}{Kunchay and
  Abdullah}{2021}]%
        {kunchay2021assessing}
\bibfield{author}{\bibinfo{person}{Sahiti Kunchay} {and} \bibinfo{person}{Saeed
  Abdullah}.} \bibinfo{year}{2021}\natexlab{}.
\newblock \showarticletitle{Assessing Effectiveness and Interpretability of
  Light Behaviors in Smart Speakers}. In \bibinfo{booktitle}{\emph{CUI 2021 -
  3rd Conference on Conversational User Interfaces}} (Bilbao (online), Spain)
  \emph{(\bibinfo{series}{CUI '21})}. \bibinfo{publisher}{Association for
  Computing Machinery}, \bibinfo{address}{New York, NY, USA}, Article
  \bibinfo{articleno}{15}, \bibinfo{numpages}{14}~pages.
\newblock
\showISBNx{9781450389983}
\urldef\tempurl%
\url{https://doi.org/10.1145/3469595.3469610}
\showDOI{\tempurl}


\bibitem[\protect\citeauthoryear{Kunchay, Wang, and Abdullah}{Kunchay
  et~al\mbox{.}}{2019}]%
        {kunchay2019investigating}
\bibfield{author}{\bibinfo{person}{Sahiti Kunchay}, \bibinfo{person}{Shan
  Wang}, {and} \bibinfo{person}{Saeed Abdullah}.}
  \bibinfo{year}{2019}\natexlab{}.
\newblock \showarticletitle{Investigating Users' Perceptions of Light Behaviors
  in Smart-Speakers}. In \bibinfo{booktitle}{\emph{Conference Companion
  Publication of the 2019 on Computer Supported Cooperative Work and Social
  Computing}} (Austin, TX, USA) \emph{(\bibinfo{series}{CSCW '19})}.
  \bibinfo{publisher}{Association for Computing Machinery},
  \bibinfo{address}{New York, NY, USA}, \bibinfo{pages}{262–266}.
\newblock
\showISBNx{9781450366922}
\urldef\tempurl%
\url{https://doi.org/10.1145/3311957.3359479}
\showDOI{\tempurl}


\bibitem[\protect\citeauthoryear{Lala, Milhorat, Inoue, Ishida, Takanashi, and
  Kawahara}{Lala et~al\mbox{.}}{2017}]%
        {lala2017attentive}
\bibfield{author}{\bibinfo{person}{Divesh Lala}, \bibinfo{person}{Pierrick
  Milhorat}, \bibinfo{person}{Koji Inoue}, \bibinfo{person}{Masanari Ishida},
  \bibinfo{person}{Katsuya Takanashi}, {and} \bibinfo{person}{Tatsuya
  Kawahara}.} \bibinfo{year}{2017}\natexlab{}.
\newblock \showarticletitle{Attentive listening system with backchanneling,
  response generation and flexible turn-taking}. In
  \bibinfo{booktitle}{\emph{Proceedings of the 18th Annual SIGdial Meeting on
  Discourse and Dialogue}}. \bibinfo{pages}{127--136}.
\newblock
\newblock
\shownote{\url{https://doi.org/10.18653/v1/W17-5516}.}


\bibitem[\protect\citeauthoryear{Lee, Yamashita, Huang, and Fu}{Lee
  et~al\mbox{.}}{2020}]%
        {lee2020hear}
\bibfield{author}{\bibinfo{person}{Yi-Chieh Lee}, \bibinfo{person}{Naomi
  Yamashita}, \bibinfo{person}{Yun Huang}, {and} \bibinfo{person}{Wai Fu}.}
  \bibinfo{year}{2020}\natexlab{}.
\newblock \showarticletitle{``I Hear You, I Feel You": Encouraging Deep
  Self-disclosure through a Chatbot}. In \bibinfo{booktitle}{\emph{Proceedings
  of the 2020 CHI conference on human factors in computing systems}}.
  \bibinfo{pages}{1--12}.
\newblock
\urldef\tempurl%
\url{https://doi.org/10.1145/3313831.3376175}
\showURL{%
\tempurl}


\bibitem[\protect\citeauthoryear{Levitt}{Levitt}{2002}]%
        {levittcounsel}
\bibfield{author}{\bibinfo{person}{Dana~Heller Levitt}.}
  \bibinfo{year}{2002}\natexlab{}.
\newblock \showarticletitle{{Active Listening and Counselor Self-Efficacy:
  Emphasis on One Microskill in Beginning Counselor Training}}.
\newblock \bibinfo{journal}{\emph{The Clinical Supervisor}}
  \bibinfo{volume}{20}, \bibinfo{number}{2} (\bibinfo{year}{2002}),
  \bibinfo{pages}{101--115}.
\newblock
\urldef\tempurl%
\url{https://doi.org/10.1300/J001v20n02_09}
\showURL{%
\tempurl}


\bibitem[\protect\citeauthoryear{Liu and Sundar}{Liu and Sundar}{2018}]%
        {liu2018should}
\bibfield{author}{\bibinfo{person}{Bingjie Liu} {and} \bibinfo{person}{S.~Shyam
  Sundar}.} \bibinfo{year}{2018}\natexlab{}.
\newblock \showarticletitle{Should Machines Express Sympathy and Empathy?
  Experiments with a Health Advice Chatbot}.
\newblock \bibinfo{journal}{\emph{Cyberpsychology, behavior and social
  networking}} \bibinfo{volume}{21}, \bibinfo{number}{10}
  (\bibinfo{year}{2018}), \bibinfo{pages}{625}.
\newblock
\newblock
\shownote{\url{https://doi.org/10.1089/cyber.2018.0110}.}


\bibitem[\protect\citeauthoryear{Lucas, Gratch, King, and Morency}{Lucas
  et~al\mbox{.}}{2014}]%
        {lucas2014s}
\bibfield{author}{\bibinfo{person}{Gale~M Lucas}, \bibinfo{person}{Jonathan
  Gratch}, \bibinfo{person}{Aisha King}, {and} \bibinfo{person}{Louis-Philippe
  Morency}.} \bibinfo{year}{2014}\natexlab{}.
\newblock \showarticletitle{It's only a computer: Virtual humans increase
  willingness to disclose}.
\newblock \bibinfo{journal}{\emph{Computers in Human Behavior}}
  \bibinfo{volume}{37} (\bibinfo{year}{2014}), \bibinfo{pages}{94--100}.
\newblock
\newblock
\shownote{\url{https://doi.org/10.1016/j.chb.2014.04.043}.}


\bibitem[\protect\citeauthoryear{Luger and Sellen}{Luger and Sellen}{2016}]%
        {10.1145/2858036.2858288}
\bibfield{author}{\bibinfo{person}{Ewa Luger} {and} \bibinfo{person}{Abigail
  Sellen}.} \bibinfo{year}{2016}\natexlab{}.
\newblock \bibinfo{booktitle}{\emph{"Like Having a Really Bad PA": The Gulf
  between User Expectation and Experience of Conversational Agents}}.
\newblock \bibinfo{publisher}{Association for Computing Machinery},
  \bibinfo{address}{New York, NY, USA}, \bibinfo{pages}{5286–5297}.
\newblock
\showISBNx{9781450333627}
\urldef\tempurl%
\url{https://doi.org/10.1145/2858036.2858288}
\showURL{%
\tempurl}


\bibitem[\protect\citeauthoryear{Miller, Berg, and Archer}{Miller
  et~al\mbox{.}}{1983}]%
        {millerOpenersIndividualsWho1983a}
\bibfield{author}{\bibinfo{person}{Lynn~C. Miller}, \bibinfo{person}{John~H.
  Berg}, {and} \bibinfo{person}{Richard~L. Archer}.}
  \bibinfo{year}{1983}\natexlab{}.
\newblock \showarticletitle{Openers: {{Individuals}} Who Elicit Intimate
  Self-Disclosure}.
\newblock \bibinfo{journal}{\emph{Journal of Personality and Social
  Psychology}} \bibinfo{volume}{44}, \bibinfo{number}{6} (\bibinfo{date}{June}
  \bibinfo{year}{1983}), \bibinfo{pages}{1234--1244}.
\newblock
\showISSN{0022-3514}
\urldef\tempurl%
\url{https://doi.org/10.1037/0022-3514.44.6.1234}
\showDOI{\tempurl}


\bibitem[\protect\citeauthoryear{Moon}{Moon}{2000}]%
        {10.1086/209566}
\bibfield{author}{\bibinfo{person}{Youngme Moon}.}
  \bibinfo{year}{2000}\natexlab{}.
\newblock \showarticletitle{{Intimate Exchanges: Using Computers to Elicit
  Self-Disclosure from Consumers}}.
\newblock \bibinfo{journal}{\emph{Journal of Consumer Research}}
  \bibinfo{volume}{26}, \bibinfo{number}{4} (\bibinfo{date}{03}
  \bibinfo{year}{2000}), \bibinfo{pages}{323--339}.
\newblock
\showISSN{0093-5301}
\urldef\tempurl%
\url{https://doi.org/10.1086/209566}
\showDOI{\tempurl}


\bibitem[\protect\citeauthoryear{Motalebi, Cho, Sundar, and Abdullah}{Motalebi
  et~al\mbox{.}}{2019}]%
        {motalebi2019}
\bibfield{author}{\bibinfo{person}{Nasim Motalebi}, \bibinfo{person}{Eugene
  Cho}, \bibinfo{person}{S.~Shyam Sundar}, {and} \bibinfo{person}{Saeed
  Abdullah}.} \bibinfo{year}{2019}\natexlab{}.
\newblock \showarticletitle{Can Alexa Be Your Therapist? How Back-Channeling
  Transforms Smart-Speakers to Be Active Listeners}. In
  \bibinfo{booktitle}{\emph{Conference Companion Publication of the 2019 on
  Computer Supported Cooperative Work and Social Computing}} (Austin, TX, USA)
  \emph{(\bibinfo{series}{CSCW '19})}. \bibinfo{publisher}{Association for
  Computing Machinery}, \bibinfo{address}{New York, NY, USA},
  \bibinfo{pages}{309–313}.
\newblock
\showISBNx{9781450366922}
\urldef\tempurl%
\url{https://doi.org/10.1145/3311957.3359502}
\showDOI{\tempurl}


\bibitem[\protect\citeauthoryear{Nass and Lee}{Nass and Lee}{2001}]%
        {nasslee2001}
\bibfield{author}{\bibinfo{person}{Clifford Nass} {and}
  \bibinfo{person}{Kwan~Min Lee}.} \bibinfo{year}{2001}\natexlab{}.
\newblock \showarticletitle{Does computer-synthesized speech manifest
  personality? Experimental tests of recognition, similarity-attraction, and
  consistency-attraction.}
\newblock  \bibinfo{volume}{7}, \bibinfo{number}{3} (\bibinfo{year}{2001}),
  \bibinfo{pages}{171–181}.
\newblock
\urldef\tempurl%
\url{https://doi.org/10.1037/1076-898X.7.3.171}
\showURL{%
\tempurl}


\bibitem[\protect\citeauthoryear{Newell and MacNeil}{Newell and
  MacNeil}{2010}]%
        {newell2010professional}
\bibfield{author}{\bibinfo{person}{Jason~M Newell} {and}
  \bibinfo{person}{Gordon~A MacNeil}.} \bibinfo{year}{2010}\natexlab{}.
\newblock \showarticletitle{Professional burnout, vicarious trauma, secondary
  traumatic stress, and compassion fatigue: A review of theoretical terms, risk
  factors, and preventive methods for clinicians and researchers}.
\newblock \bibinfo{journal}{\emph{Best Practices in Mental Health: An
  International Journal}} (\bibinfo{year}{2010}).
\newblock


\bibitem[\protect\citeauthoryear{Nicholson}{Nicholson}{2012}]%
        {nicholson2012review}
\bibfield{author}{\bibinfo{person}{Nicholas~R Nicholson}.}
  \bibinfo{year}{2012}\natexlab{}.
\newblock \showarticletitle{A review of social isolation: an important but
  underassessed condition in older adults}.
\newblock \bibinfo{journal}{\emph{The journal of primary prevention}}
  \bibinfo{volume}{33}, \bibinfo{number}{2-3} (\bibinfo{year}{2012}),
  \bibinfo{pages}{137--152}.
\newblock
\newblock
\shownote{\url{https://doi.org/10.1007/s10935-012-0271-2}.}


\bibitem[\protect\citeauthoryear{Nils and Rim{\'e}}{Nils and Rim{\'e}}{2012}]%
        {nils2012beyond}
\bibfield{author}{\bibinfo{person}{Fr{\'e}d{\'e}ric Nils} {and}
  \bibinfo{person}{Bernard Rim{\'e}}.} \bibinfo{year}{2012}\natexlab{}.
\newblock \showarticletitle{Beyond the myth of venting: Social sharing modes
  determine the benefits of emotional disclosure}.
\newblock \bibinfo{journal}{\emph{European Journal of Social Psychology}}
  \bibinfo{volume}{42}, \bibinfo{number}{6} (\bibinfo{year}{2012}),
  \bibinfo{pages}{672--681}.
\newblock
\newblock
\shownote{\url{https://doi.org/10.1002/ejsp.1880}.}


\bibitem[\protect\citeauthoryear{Oertel, Lopes, Yu, Mora, Gustafson, Black, and
  Odobez}{Oertel et~al\mbox{.}}{2016}]%
        {oertel2016towards}
\bibfield{author}{\bibinfo{person}{Catharine Oertel}, \bibinfo{person}{Jos{\'e}
  Lopes}, \bibinfo{person}{Yu Yu}, \bibinfo{person}{Kenneth A~Funes Mora},
  \bibinfo{person}{Joakim Gustafson}, \bibinfo{person}{Alan~W Black}, {and}
  \bibinfo{person}{Jean-Marc Odobez}.} \bibinfo{year}{2016}\natexlab{}.
\newblock \showarticletitle{Towards building an attentive artificial listener:
  On the perception of attentiveness in audio-visual feedback tokens}. In
  \bibinfo{booktitle}{\emph{Proceedings of the 18th ACM International
  Conference on Multimodal Interaction}}. \bibinfo{pages}{21--28}.
\newblock
\newblock
\shownote{\url{https://doi.org/10.1145/2993148.2993188}.}


\bibitem[\protect\citeauthoryear{Pennebaker}{Pennebaker}{1997}]%
        {pennebaker1997writing}
\bibfield{author}{\bibinfo{person}{James~W Pennebaker}.}
  \bibinfo{year}{1997}\natexlab{}.
\newblock \showarticletitle{Writing about emotional experiences as a
  therapeutic process}.
\newblock \bibinfo{journal}{\emph{Psychological science}} \bibinfo{volume}{8},
  \bibinfo{number}{3} (\bibinfo{year}{1997}), \bibinfo{pages}{162--166}.
\newblock
\newblock
\shownote{\url{https://doi.org/10.1111/j.1467-9280.1997.tb00403.x}.}


\bibitem[\protect\citeauthoryear{Poppe~R.}{Poppe~R.}{2011}]%
        {TimingMatters}
\bibfield{author}{\bibinfo{person}{Heylen~D. Poppe~R., Truong~K.P.}}
  \bibinfo{year}{2011}\natexlab{}.
\newblock \showarticletitle{Backchannels: quantity, type and timing matters}.
\newblock \bibinfo{journal}{\emph{Intelligent Virtual Agents. IVA 2011}}
  \bibinfo{volume}{6859} (\bibinfo{year}{2011}), \bibinfo{pages}{228--239}.
\newblock


\bibitem[\protect\citeauthoryear{Porcheron, Clark, Jones, Candello, Cowan,
  Murad, Sin, Aylett, Lee, Munteanu, Fischer, Doyle, and Kaye}{Porcheron
  et~al\mbox{.}}{2020}]%
        {10.1145/3406865.3418587}
\bibfield{author}{\bibinfo{person}{Martin Porcheron}, \bibinfo{person}{Leigh
  Clark}, \bibinfo{person}{Matt Jones}, \bibinfo{person}{Heloisa Candello},
  \bibinfo{person}{Benjamin~R. Cowan}, \bibinfo{person}{Christine Murad},
  \bibinfo{person}{Jaisie Sin}, \bibinfo{person}{Matthew~P. Aylett},
  \bibinfo{person}{Minha Lee}, \bibinfo{person}{Cosmin Munteanu},
  \bibinfo{person}{Joel~E. Fischer}, \bibinfo{person}{Philip~R. Doyle}, {and}
  \bibinfo{person}{Jofish Kaye}.} \bibinfo{year}{2020}\natexlab{}.
\newblock \bibinfo{booktitle}{\emph{CUI@CSCW: Collaborating through
  Conversational User Interfaces}}.
\newblock \bibinfo{publisher}{Association for Computing Machinery},
  \bibinfo{address}{New York, NY, USA}, \bibinfo{pages}{483–492}.
\newblock
\showISBNx{9781450380591}
\urldef\tempurl%
\url{https://doi.org/10.1145/3406865.3418587}
\showURL{%
\tempurl}


\bibitem[\protect\citeauthoryear{Porcheron, Fischer, McGregor, Brown, Luger,
  Candello, and O'Hara}{Porcheron et~al\mbox{.}}{2017b}]%
        {porcheron2017talking}
\bibfield{author}{\bibinfo{person}{Martin Porcheron}, \bibinfo{person}{Joel~E
  Fischer}, \bibinfo{person}{Moira McGregor}, \bibinfo{person}{Barry Brown},
  \bibinfo{person}{Ewa Luger}, \bibinfo{person}{Heloisa Candello}, {and}
  \bibinfo{person}{Kenton O'Hara}.} \bibinfo{year}{2017}\natexlab{b}.
\newblock \showarticletitle{Talking with conversational agents in collaborative
  action}. In \bibinfo{booktitle}{\emph{companion of the 2017 ACM conference on
  computer supported cooperative work and social computing}}.
  \bibinfo{pages}{431--436}.
\newblock
\urldef\tempurl%
\url{https://doi.org/10.1145/3022198.3022666}
\showDOI{\tempurl}


\bibitem[\protect\citeauthoryear{Porcheron, Fischer, and Sharples}{Porcheron
  et~al\mbox{.}}{2017a}]%
        {porcheron2017animals}
\bibfield{author}{\bibinfo{person}{Martin Porcheron}, \bibinfo{person}{Joel~E
  Fischer}, {and} \bibinfo{person}{Sarah Sharples}.}
  \bibinfo{year}{2017}\natexlab{a}.
\newblock \showarticletitle{``Do Animals Have Accents?'' Talking with Agents in
  Multi-Party Conversation}. In \bibinfo{booktitle}{\emph{Proceedings of the
  2017 ACM conference on computer supported cooperative work and social
  computing}}. \bibinfo{pages}{207--219}.
\newblock
\urldef\tempurl%
\url{https://doi.org/10.1145/2998181.2998298}
\showDOI{\tempurl}


\bibitem[\protect\citeauthoryear{Pradhan, Lazar, and Findlater}{Pradhan
  et~al\mbox{.}}{2020}]%
        {10.1145/3373759}
\bibfield{author}{\bibinfo{person}{Alisha Pradhan}, \bibinfo{person}{Amanda
  Lazar}, {and} \bibinfo{person}{Leah Findlater}.}
  \bibinfo{year}{2020}\natexlab{}.
\newblock \showarticletitle{Use of Intelligent Voice Assistants by Older Adults
  with Low Technology Use}.
\newblock \bibinfo{journal}{\emph{ACM Trans. Comput.-Hum. Interact.}}
  \bibinfo{volume}{27}, \bibinfo{number}{4}, Article \bibinfo{articleno}{31}
  (\bibinfo{date}{Sept.} \bibinfo{year}{2020}), \bibinfo{numpages}{27}~pages.
\newblock
\showISSN{1073-0516}
\urldef\tempurl%
\url{https://doi.org/10.1145/3373759}
\showDOI{\tempurl}


\bibitem[\protect\citeauthoryear{Qiu, Kanski, Doerksen, Winkels, Schmitz, and
  Abdullah}{Qiu et~al\mbox{.}}{2021}]%
        {qiu2021nurse}
\bibfield{author}{\bibinfo{person}{Ling Qiu}, \bibinfo{person}{Bethany Kanski},
  \bibinfo{person}{Shawna Doerksen}, \bibinfo{person}{Renate Winkels},
  \bibinfo{person}{Kathryn~H Schmitz}, {and} \bibinfo{person}{Saeed Abdullah}.}
  \bibinfo{year}{2021}\natexlab{}.
\newblock \bibinfo{booktitle}{\emph{Nurse AMIE: Using Smart Speakers to Provide
  Supportive Care Intervention for Women with Metastatic Breast Cancer}}.
\newblock \bibinfo{publisher}{Association for Computing Machinery},
  \bibinfo{address}{New York, NY, USA}.
\newblock
\showISBNx{9781450380959}
\urldef\tempurl%
\url{https://doi.org/10.1145/3411763.3451827}
\showURL{%
\tempurl}


\bibitem[\protect\citeauthoryear{Rauvola, Vega, and Lavigne}{Rauvola
  et~al\mbox{.}}{2019}]%
        {rauvola_compassion_2019}
\bibfield{author}{\bibinfo{person}{Rachel~S. Rauvola},
  \bibinfo{person}{Dulce~M. Vega}, {and} \bibinfo{person}{Kristi~N. Lavigne}.}
  \bibinfo{year}{2019}\natexlab{}.
\newblock \showarticletitle{Compassion {Fatigue}, {Secondary} {Traumatic}
  {Stress}, and {Vicarious} {Traumatization}: a {Qualitative} {Review} and
  {Research} {Agenda}}.
\newblock \bibinfo{journal}{\emph{Occupational Health Science}}
  \bibinfo{volume}{3}, \bibinfo{number}{3} (\bibinfo{date}{Sept.}
  \bibinfo{year}{2019}), \bibinfo{pages}{297--336}.
\newblock
\showISSN{2367-0142}
\urldef\tempurl%
\url{https://doi.org/10.1007/s41542-019-00045-1}
\showDOI{\tempurl}


\bibitem[\protect\citeauthoryear{Rim{\'e}}{Rim{\'e}}{2009}]%
        {rime2009emotion}
\bibfield{author}{\bibinfo{person}{Bernard Rim{\'e}}.}
  \bibinfo{year}{2009}\natexlab{}.
\newblock \showarticletitle{Emotion elicits the social sharing of emotion:
  Theory and empirical review}.
\newblock \bibinfo{journal}{\emph{Emotion review}} \bibinfo{volume}{1},
  \bibinfo{number}{1} (\bibinfo{year}{2009}), \bibinfo{pages}{60--85}.
\newblock
\newblock
\shownote{\url{https://doi.org/10.1177/1754073908097189}.}


\bibitem[\protect\citeauthoryear{Rimé, Finkenauer, Luminet, Zech, and
  Philippot}{Rimé et~al\mbox{.}}{1998}]%
        {rime_social_1998}
\bibfield{author}{\bibinfo{person}{Bernard Rimé}, \bibinfo{person}{Catrin
  Finkenauer}, \bibinfo{person}{Olivier Luminet}, \bibinfo{person}{Emmanuelle
  Zech}, {and} \bibinfo{person}{Pierre Philippot}.}
  \bibinfo{year}{1998}\natexlab{}.
\newblock \showarticletitle{Social {Sharing} of {Emotion}: {New} {Evidence} and
  {New} {Questions}}.
\newblock \bibinfo{journal}{\emph{European Review of Social Psychology}}
  \bibinfo{volume}{9}, \bibinfo{number}{1} (\bibinfo{date}{Jan.}
  \bibinfo{year}{1998}), \bibinfo{pages}{145--189}.
\newblock
\showISSN{1046-3283, 1479-277X}
\urldef\tempurl%
\url{https://doi.org/10.1080/14792779843000072}
\showDOI{\tempurl}


\bibitem[\protect\citeauthoryear{Rimé, Mesquita, Boca, and Philippot}{Rimé
  et~al\mbox{.}}{1991}]%
        {rime1991beyond}
\bibfield{author}{\bibinfo{person}{Bernard Rimé}, \bibinfo{person}{Batja
  Mesquita}, \bibinfo{person}{Stefano Boca}, {and} \bibinfo{person}{Pierre
  Philippot}.} \bibinfo{year}{1991}\natexlab{}.
\newblock \showarticletitle{Beyond the emotional event: {Six} studies on the
  social sharing of emotion}.
\newblock \bibinfo{journal}{\emph{Cognition \& Emotion}} \bibinfo{volume}{5},
  \bibinfo{number}{5-6} (\bibinfo{date}{Sept.} \bibinfo{year}{1991}),
  \bibinfo{pages}{435--465}.
\newblock
\showISSN{0269-9931, 1464-0600}
\urldef\tempurl%
\url{https://doi.org/10.1080/02699939108411052}
\showDOI{\tempurl}


\bibitem[\protect\citeauthoryear{Robertson}{Robertson}{2005}]%
        {robertson_active_2005}
\bibfield{author}{\bibinfo{person}{Kathryn Robertson}.}
  \bibinfo{year}{2005}\natexlab{}.
\newblock \showarticletitle{Active listening: more than just paying attention}.
\newblock \bibinfo{journal}{\emph{Australian Family Physician}}
  \bibinfo{volume}{34}, \bibinfo{number}{12} (\bibinfo{year}{2005}),
  \bibinfo{pages}{1053--1055}.
\newblock
\showISSN{0300-8495}


\bibitem[\protect\citeauthoryear{Rost and Candlin}{Rost and Candlin}{2014}]%
        {rost2014listening}
\bibfield{author}{\bibinfo{person}{Michael Rost} {and} \bibinfo{person}{CN
  Candlin}.} \bibinfo{year}{2014}\natexlab{}.
\newblock \bibinfo{booktitle}{\emph{Listening in language learning}}.
\newblock \bibinfo{publisher}{Routledge}.
\newblock
\newblock
\shownote{\url{https://doi.org/10.4324/9781315846699}.}


\bibitem[\protect\citeauthoryear{Sannon, Stoll, DiFranzo, Jung, and
  Bazarova}{Sannon et~al\mbox{.}}{2018}]%
        {sannon2018personification}
\bibfield{author}{\bibinfo{person}{Shruti Sannon}, \bibinfo{person}{Brett
  Stoll}, \bibinfo{person}{Dominic DiFranzo}, \bibinfo{person}{Malte Jung},
  {and} \bibinfo{person}{Natalya~N Bazarova}.} \bibinfo{year}{2018}\natexlab{}.
\newblock \showarticletitle{How personification and interactivity influence
  stress-related disclosures to conversational agents}. In
  \bibinfo{booktitle}{\emph{companion of the 2018 ACM conference on computer
  supported cooperative work and social computing}}. \bibinfo{pages}{285--288}.
\newblock
\urldef\tempurl%
\url{https://doi.org/10.1145/3272973.3274076}
\showDOI{\tempurl}


\bibitem[\protect\citeauthoryear{Sciuto, Saini, Forlizzi, and Hong}{Sciuto
  et~al\mbox{.}}{2018}]%
        {10.1145/3196709.3196772}
\bibfield{author}{\bibinfo{person}{Alex Sciuto}, \bibinfo{person}{Arnita
  Saini}, \bibinfo{person}{Jodi Forlizzi}, {and} \bibinfo{person}{Jason~I.
  Hong}.} \bibinfo{year}{2018}\natexlab{}.
\newblock \showarticletitle{"Hey Alexa, What's Up?": A Mixed-Methods Studies of
  In-Home Conversational Agent Usage}. In \bibinfo{booktitle}{\emph{Proceedings
  of the 2018 Designing Interactive Systems Conference}} (Hong Kong, China)
  \emph{(\bibinfo{series}{DIS '18})}. \bibinfo{publisher}{Association for
  Computing Machinery}, \bibinfo{address}{New York, NY, USA},
  \bibinfo{pages}{857–868}.
\newblock
\showISBNx{9781450351980}
\urldef\tempurl%
\url{https://doi.org/10.1145/3196709.3196772}
\showDOI{\tempurl}


\bibitem[\protect\citeauthoryear{Sloan and Marx}{Sloan and Marx}{2004}]%
        {sloan2004taking}
\bibfield{author}{\bibinfo{person}{Denise~M Sloan} {and}
  \bibinfo{person}{Brian~P Marx}.} \bibinfo{year}{2004}\natexlab{}.
\newblock \showarticletitle{Taking pen to hand: Evaluating theories underlying
  the written disclosure paradigm}.
\newblock \bibinfo{journal}{\emph{Clinical psychology: Science and practice}}
  \bibinfo{volume}{11}, \bibinfo{number}{2} (\bibinfo{year}{2004}),
  \bibinfo{pages}{121--137}.
\newblock
\newblock
\shownote{\url{https://doi.org/10.1093/clipsy.bph062}.}


\bibitem[\protect\citeauthoryear{Sundar, Xu, Bellur, Oh, and Jia}{Sundar
  et~al\mbox{.}}{2011}]%
        {sundar2011beyond}
\bibfield{author}{\bibinfo{person}{S.~Shyam Sundar}, \bibinfo{person}{Qian Xu},
  \bibinfo{person}{Saraswathi Bellur}, \bibinfo{person}{Jeeyun Oh}, {and}
  \bibinfo{person}{Haiyan Jia}.} \bibinfo{year}{2011}\natexlab{}.
\newblock \showarticletitle{Beyond pointing and clicking: how do newer
  interaction modalities affect user engagement?}
\newblock In \bibinfo{booktitle}{\emph{CHI'11 Extended Abstracts on Human
  Factors in Computing Systems}}. \bibinfo{pages}{1477--1482}.
\newblock
\newblock
\shownote{\url{https://doi.org/10.1145/1979742.1979794}.}


\bibitem[\protect\citeauthoryear{Tausczik and Pennebaker}{Tausczik and
  Pennebaker}{2010}]%
        {tausczikPsychologicalMeaningWords2010}
\bibfield{author}{\bibinfo{person}{Yla~R. Tausczik} {and}
  \bibinfo{person}{James~W. Pennebaker}.} \bibinfo{year}{2010}\natexlab{}.
\newblock \showarticletitle{The {{Psychological Meaning}} of {{Words}}:
  {{LIWC}} and {{Computerized Text Analysis Methods}}}.
\newblock \bibinfo{journal}{\emph{Journal of Language and Social Psychology}}
  \bibinfo{volume}{29}, \bibinfo{number}{1} (\bibinfo{date}{March}
  \bibinfo{year}{2010}), \bibinfo{pages}{24--54}.
\newblock
\showISSN{0261-927X, 1552-6526}
\urldef\tempurl%
\url{https://doi.org/10.1177/0261927X09351676}
\showDOI{\tempurl}


\bibitem[\protect\citeauthoryear{Ward}{Ward}{1996}]%
        {Ward1996UsingPC}
\bibfield{author}{\bibinfo{person}{Nigel~G. Ward}.}
  \bibinfo{year}{1996}\natexlab{}.
\newblock \showarticletitle{Using prosodic clues to decide when to produce
  back-channel utterances}.
\newblock \bibinfo{journal}{\emph{Proceeding of Fourth International Conference
  on Spoken Language Processing. ICSLP '96}}  \bibinfo{volume}{3}
  (\bibinfo{year}{1996}), \bibinfo{pages}{1728--1731}.
\newblock


\bibitem[\protect\citeauthoryear{Watson, Clark, and Tellegen}{Watson
  et~al\mbox{.}}{1988}]%
        {watson1988development}
\bibfield{author}{\bibinfo{person}{David Watson}, \bibinfo{person}{Lee~Anna
  Clark}, {and} \bibinfo{person}{Auke Tellegen}.}
  \bibinfo{year}{1988}\natexlab{}.
\newblock \showarticletitle{Development and validation of brief measures of
  positive and negative affect: the PANAS scales}.
\newblock \bibinfo{journal}{\emph{Journal of personality and social
  psychology}} \bibinfo{volume}{54}, \bibinfo{number}{6}
  (\bibinfo{year}{1988}), \bibinfo{pages}{1063}.
\newblock
\newblock
\shownote{\url{https://doi.org/10.1037/0022-3514.54.6.1063}.}


\bibitem[\protect\citeauthoryear{Weger, Castle~Bell, Minei, and Robinson}{Weger
  et~al\mbox{.}}{2014}]%
        {wegerRelativeEffectivenessActive2014}
\bibfield{author}{\bibinfo{person}{Harry Weger}, \bibinfo{person}{Gina
  Castle~Bell}, \bibinfo{person}{Elizabeth~M. Minei}, {and}
  \bibinfo{person}{Melissa~C. Robinson}.} \bibinfo{year}{2014}\natexlab{}.
\newblock \showarticletitle{The {{Relative Effectiveness}} of {{Active
  Listening}} in {{Initial Interactions}}}.
\newblock \bibinfo{journal}{\emph{International Journal of Listening}}
  \bibinfo{volume}{28}, \bibinfo{number}{1} (\bibinfo{date}{Jan.}
  \bibinfo{year}{2014}), \bibinfo{pages}{13--31}.
\newblock
\showISSN{1090-4018, 1932-586X}
\urldef\tempurl%
\url{https://doi.org/10.1080/10904018.2013.813234}
\showDOI{\tempurl}


\bibitem[\protect\citeauthoryear{Weizenbaum}{Weizenbaum}{1966}]%
        {Eliza}
\bibfield{author}{\bibinfo{person}{Joseph Weizenbaum}.}
  \bibinfo{year}{1966}\natexlab{}.
\newblock \showarticletitle{ELIZA—a Computer Program for the Study of Natural
  Language Communication between Man and Machine}.
\newblock \bibinfo{journal}{\emph{Commun. ACM}} \bibinfo{volume}{9},
  \bibinfo{number}{1} (\bibinfo{date}{Jan.} \bibinfo{year}{1966}),
  \bibinfo{pages}{36–45}.
\newblock
\showISSN{0001-0782}
\urldef\tempurl%
\url{https://doi.org/10.1145/365153.365168}
\showDOI{\tempurl}


\end{thebibliography}

\end{document}